\documentclass[onecolumn,authoryear]{els-mrw} 

\usepackage{amsmath,amssymb,amsfonts,amsthm,makeidx,graphicx}
\usepackage{txfonts}
\usepackage{helvet}
\usepackage{ulem}
\usepackage{hyperref}
\usepackage{minitoc}
\newcommand{\grl}{Geophysical Research Letters}
\newcommand{\aap}{Astronomy \& Astrophysics}
\newcommand{\apj}{The Astrophysical Journal}
\newcommand{\apjs}{The Astrophysical Journal Supplement}
\newcommand{\prb}{Physical Review B}

\newcommand{\prl}{Physical Review Letters}
\newcommand{\icarus}{Icarus}
\newcommand{\mnras}{Monthly Notices of the Royal Astronomical Society}
\newcommand{\planss}{Planetary and Space Science}
\newcommand{\psj}{The Planetary Science Journal}
\newcommand{\ssr}{Space Science Reviews}

\begin{document}
\dominitoc
\chapter{Giant planet  interiors and atmospheres}\label{chap1}

\author[1]{Ravit Helled}%
\author[1]{Saburo Howard}%

\address[1]{\orgname{Department of Astrophysics}, \orgdiv{University of Zurich}, \orgaddress{Winterthurerstr. 190, 8057 Zurich, Switzerland}}


\articletag{Chapter Article tagline: update of previous edition,, reprint..}

\maketitle

\begin{abstract}[Abstract]
 Studying the interiors of the outer planets is crucial for a comprehensive understanding of our planetary system, and provides key knowledge on the origin of the solar system, the behavior of materials at extreme conditions, the relation between the planetary interior and atmosphere, and exoplanet characterization. 
In this review, we summarize the current understanding of the interiors and atmospheres of the giant planets in the solar system: Jupiter, Saturn, Uranus and Neptune. 
We describe the principles of interior structure and evolution models and discuss the link between the planetary atmospheres and deep interiors. We summarize the current understanding of the bulk compositions and internal structures of the outer planets and the remaining future challenges. 
\end{abstract}

\subsection*{Keywords:} Jupiter, Saturn, Uranus, Neptune, giant planets, ice giants, gaseous planets, interior models, planetary structure


\minitoc

\section*{Learning Objectives:}
By the end of this chapter, you should understand: 
\begin{itemize}%
\item  How to build an interior model 
\item What measurements are used to model the internal structures of giant planets
\item What are the main characteristics of Jupiter, Saturn, Uranus and Neptune
\item What are the key open questions related to the interiors of the outer planets
\end{itemize}

\section*{Glossary:}
{\bf Adiabatic gradient} The temperature gradient in a region of constant composition and entropy.\\
{\bf Bond albedo} The fraction of light that is reflected back to space by the planet.\\
{\bf Convection} A heat  transport mechanism where heat is transferred by movements of a fluid.\\
{\bf Core} The central region of a planet which is composed of pure heavy-elements.\\
{\bf Density functional theory (DFT)} A computational quantum mechanical modelling method used to study the electronic structure of many-body systems. \\
{\bf Equation of state (EoS)} An equation (or a numerical table) that relates the values of pressure, volume, and temperature of a given substance in thermodynamic equilibrium. \\
{\bf Fuzzy core} The central region of the planet but with the heavy elements being gradually mixed with hydrogen and helium (the envelope). The term "compositional gradient" is also frequently used.\\
{\bf Gaseous envelope} The region in giant planets' interiors where the composition is dominated by hydrogen and helium.\\
{\bf Gravitational moments} Coefficients used to  describe the nonuniform distribution of mass (density) in a planet.\\
{\bf Heavy elements} All the elements in the periodic table that are heavier than helium.\\
{\bf Quantum Monte Carlo (QMC)} A computational method to  study complex quantum systems. This method provides an accurate approximation of the quantum many-body problem.\\
{\bf Luminosity} An absolute measure of the radiant power of an astronomical object.\\
{\bf Markov chain Monte Carlo (MCMC)}  A class of algorithms used to sample models based on given data. \\
{\bf Mass fraction} The ratio of the mass of a constituent to the total mass of the mixture. \\
{\bf Metallicity} The mass fraction of heavy elements.\\
{\bf Moment of inertia} A measure of the rotational inertia of an object around a specific axis of rotation.\\
{\bf Phase diagram} A graph showing the limiting conditions for solid, liquid, and gaseous phases of a single substance or of a mixture of substances. \\
{\bf Phase separation} A process in which a mixture separates into two or more distinct phases. The term "demixing" is also frequently used. \\
{\bf Proto-planetary disk} The disk that surrounds a young star where planets can form. The disk is composed mostly of gas (hydrogen and helium) and some dust (of the order of a percent).\\
{\bf Proto-solar composition} The composition of the nebula that formed the Solar System.\\ 
{\bf Rosseland mean opacity} The total amount of energy absorbed over all wavelengths. \\
{\bf Surface (of a giant planet)} The location where the pressure is 1 bar, similar to the pressure on Earth's surface.


\section{Introduction: Why study the interiors of giant planets?}\label{chap1:sec1}

Giant planets are key planets to investigate because they play a crucial role in shaping the architecture of young planetary systems. This is due to their fast formation - they must form early enough to accrete hydrogen-helium gas from the proto-planetary disks, the birth places of planets, and their large gravity influences the dynamics of other objects. In addition, the composition of giant planets provides information on the physical and chemical properties of the proto-planetary disks which led to the formation of the solar system's planets. Finally, giant planets are natural laboratories for investigating the behavior of materials  at extreme high pressures and temperatures that are challenging to replicate and measure on Earth. 
\par 

Despite decades of investigations on both the theoretical and observational fronts, we still do not have unique and self-consistent views of the interiors of Jupiter, Saturn, Uranus and Neptune. 
However, significant progress has been made thanks to new accurate measurements from spacecraft that led to an improved  understanding of the relevant physical/chemical processes that need to be included in models. In addition, now we can clearly identify knowledge gaps and required measurements from future missions. Below we summarize the current knowledge of the internal structures of the gas giants, Jupiter, Saturn, and the ice giants, Uranus and Neptune. 

\section{How to make an interior model?}
\label{sec:make_model}
Knowledge on the internal structure of the outer planets in the solar system must be inferred from models that fit their observed physical properties.  Key measurements for structure models include the planetary mass, radius, gravitational
fields, 1-bar temperatures, atmospheric
composition, and rotation rates.  The fundamental physical properties of the four planets are listed in Table \ref{tab:properties_planets}. 
Several space missions have visited the outer planets in the solar system. Jupiter has been explored by multiple missions: NASA's Pioneer 10 and 11 were the first to provide close-up images and data, followed by the Voyager 1 and 2 flybys that revealed intricate details about its moons and rings. The Galileo orbiter provided extensive data from 1995 to 2003, and the Juno mission, launched in 2011, is still orbiting Jupiter as this review is being written in 2024. The Juno mission provided accurate measurements of Jupiter's atmospheric composition and gravitational and magnetic fields. Saturn was first visited by Pioneer 11 and later by Voyager 1 and 2 where  the most detailed exploration came from the Cassini-Huygens mission (1997-2017), which delivered in-depth information about Saturn and its moons. Uranus and Neptune were both visited by Voyager 2 in the 1980s, which remains the only mission to have flown by these planets. As  many open questions regarding the nature of these planets remain, future missions to study Uranus and Neptune are foreseen and have been found to be a high priority by the recent planetary decadal survey (\url{https://nap.nationalacademies.org/catalog/26522/origins-worlds-and-life-a-decadal-strategy-for-planetary-science}). 
\par  

\begin{table}[h!]
{\small 
\def\arraystretch{1.2}
\centering 
\begin{tabular}{lllll}
\hline
\hline
{\bf Parameter} & {\bf Jupiter} & {\bf Saturn} & {\bf Uranus} & {\bf Neptune}\\
\hline
Semi-major axis$^a$ (AU) &  5.20336301   &  9.53707032 & 19.19126393  & 30.06896348   \\
Mass ($10^{24}$ kg)  & 1898.125 ±0.088 & 568.317 ±0.026 & 86.8099 ±0.0040  & 102.4092 ±0.0048 \\
Mean Radius (km)& 69911 ±6 & 58232 ±6 & 25362 ±7 & 24622 ±19  \\
Equatorial Radius (km) & 71492 ±4  & 60268 ±4 & 25559 ±4  &24764 ±15 \\
Mean Density (g cm$^{-3}$)$^b$ & $1.3262\pm{}0.0004$ & $0.6871\pm{}0.0002$ & 1.270 $\pm$ 0.001 & 1.638 $\pm$ 0.004 \\
R$_{\rm ref}$$^c$ (km) & 71,492 & 60,330 & 25,559 & 24,764 \\
J$_2$ ($\times$10$^{6}$) & 14,696.5735 $\pm$ 0.00056 & 16,290.573 $\pm$ 0.0093  &3510.68 $\pm$ 0.70& 3408.43 $\pm$ 4.50 \\
J$_4$ ($\times$10$^{6}$) & $-$586.6085 $\pm$ 0.0008 & $-$935.314 $\pm$ 0.0123 & $-$34.17 $\pm$ 1.30 & $-$33.40 $\pm$ 2.90 \\
J$_6$ ($\times$10$^{6}$) & 34.2007 $\pm$ 0.00223 & 86.340 $\pm$ 0.029 & 46.12--59.90$^d$  & 45.26--68.63$^d$ \\
J$_8$ ($\times$10$^{6}$) & $-$2.422 $\pm$ 0.007 & $-$14.624 $\pm$ 0.0683 & -8.4 to -17.8$^d$& -7.85 to  -20.16$^d$\\
Moment of inertia (I/MR$^2$) & 0.254 & 0.210 & 0.225 & 0.23-0.25$^d$ \\
Rotation period$^d$ (hr) & 9.9250 & 10.656 & 17.24  & 16.11 \\
1-bar Temperature (K) & $165 \pm 5^f$  & $135 \pm 5$ & 76 $\pm$ 2& 72 $\pm$ 2\\
Emitted power ($10^{16}~$J/s) & 83.65 $\pm$ 0.84 & $19.77 \pm 0.32$ & 0.560 $\pm$ 0.011& 0.534 $\pm$ 0.029\\
Absorbed power ($10^{16}~$J/s) & 50.14 $\pm$ 2.48 & $11.14 \pm 0.50$ & 0.526 $\pm$ 0.037& 0.204 $\pm$ 0.019\\
Bond Albedo $A$ & 0.343 & 0.342 & 0.300 & 0.290 \\
\hline
\hline
\end{tabular}
\caption{Basic physical properties of Jupiter, Saturn, Uranus and Neptune (from \url{https://nssdc.gsfc.nasa.gov/planetary/} and \url{https://ssd.jpl.nasa.gov/planets/phys_par.html}). $^a$J2000. $^b$The bulk Density computed based on the volume of a sphere with the published mean radius. $^c$R$_{\rm ref}$ is the reference equatorial radius in respect to the measured gravitational harmonics. $^d$Calculated values from \citep{Neuenschwander2022}. 
$^e$Note that the rotation periods of the planets are not well determined, in particular in the case of Uranus and Neptune \citep{Helled2010shape}. $^f$The value from Voyager is displayed. 
}}
\label{tab:properties_planets}       
\end{table}

Interior models are complex, and yet, they are still based on several simplifying assumptions. For example, interior models are typically one/two dimensional, spherically symmetric and are  hydrostatic (i.e., do not include dynamics). Unlike the terrestrial planets, all the outer planets in the solar system have no solid surface and therefore the "surface" of the planet is defined as the location where the pressure is 1 bar, similar to the pressure on Earth's surface. This represents the outer boundary condition for the models, and the deep interior is then modeled to the planetary center. Constraining the planetary interior relies on  measurements of the planet's gravitational field. 
The total potential of a giant planet is the sum of the gravitational and centrifugal potentials, and  is given by: 
\begin{equation}
    \begin{split}
    U(r, \theta, \phi) &= \frac{GM}{r}\Biggl(1- \sum_{n = 1}^{\infty}{{\Bigl(\frac{a}{r}\Bigr)}^{2n}}J_{2n}P_{2n}(cos\theta)\Biggr)+\frac{1}{2}\omega^2 r^2 sin^2\theta,  \\
    \label{grav_field}
    \end{split}
\end{equation}
where $G$ is the gravitational constant, $M$ is the total planetary mass, $a$ is the equatorial radius at 1 bar and $\omega$ is the angular velocity of rotation. $J_{2n}$ are the gravitational moments and $P_{2n}$ are the Legendre polynomials. $r$ and $\theta$ are the radius and the co-latitude, respectively. 
When assuming a hydrostatic planet that is spherically symmetric, only the even gravitational coefficients contribute to the gravitational field. The gravitational moments are measured by tracking the trajectory of a spacecraft orbiting (or flying by) the planet and are used to constrain the planetary density profile $\rho(r,\theta)$, via: 
\begin{equation}
    J_{2n}=-\frac{1}{Ma^{2n}}\int \rho(r',\theta)(r')^{2n}P_{2n}(\textrm{cos}\theta)d\tau',
\end{equation} 
where $d\tau'$ is a volume element and the integrals are evaluated over the planet's volume.
Interior models are set to find density profiles that fit all the observational constraints using the  
structure equations which include the mass conservation, hydrostatic balance, and thermodynamic equations as follows:
\begin{subequations}\label{eqs:struct}
\begin{align}
\frac{dm}{dr} &= 4\pi{}r^2\rho, \\
\label{eq:HE_omega}\frac{1}{\rho}\frac{dP}{dr} &= -\frac{Gm}{r^2}
+ \frac{2}{3}\omega^2r,\\
\label{eq:T}\frac{dT}{dr} &= \frac{T}{P}\frac{dP}{dr}\nabla_T,
\end{align}
\end{subequations}
where $P$ is the pressure, $T$ is the temperature, $\rho$ is the density, and $m$ is the mass. The first
equation defines the transformation between a mass variable and radius variable.
The second equation is the condition of hydrostatic equilibrium, including a
correction term to account for planetary flattening  due to uniform rotation. However, while most of the interior rotates uniformly, the winds in the atmosphere rotate differentially and can extend, in the cases of Jupiter and Saturn, down to $\sim 3000$ and $\sim 9000~$km respectively. This outer part of the planet rotates along cylinders and leads to density anomalies, thereby contributing to the gravitational moments. In Uranus and Neptune the winds are expected to penetrate to depth of the order of $\sim 1000~$km. The third
equation describes the energy transport outward from the interior of the object
to its surface. The temperature gradient, $\nabla_T\equiv{}d\ln{T}/d\ln{P}$ 
depends on how energy is transported. 
In case of convection, the temperature
gradient is set to the adiabatic gradient
$\nabla_{\rm ad}=\frac{d\ln{T}}{d\ln{P}}\arrowvert_S$, where $S$ is the
entropy. In the case of radiation the diffusion
approximation is used and the temperature gradient is given by: 
\begin{equation}
\nabla_T = \nabla_{\rm rad} = \frac{3}{16\pi{}Gac}
\frac{\kappa_RLP}{mT^4}, 
\label{eq:radg}
\end{equation}
where $\kappa_R$ is the Rosseland mean
opacity, $c$ is the speed of light and $L$ is the luminosity. Typically, radiation and conduction are combined and  an effective
opacity that accounts for the contribution of both the radiation and conductive opacities is used. 
In order to model the planet a fourth
equation is needed which relates the density, pressure, and
temperature. This is known as an \emph{equation of state} (EoS). While simple EoSs  such as the ideal gas equation exist, when it comes to planetary interiors  the EoS is very complex since it includes the behavior of hydrogen, helium and their interaction as well as the other materials that exist in the planetary interior, such as water and rocks. 


\subsubsection{Equations of state}
\label{sec:eos}
Investigating the EoS of materials at a large range of pressures and temperatures is a research field by itself. Below, we briefly summarize the relevant EoSs relevant for  studying the interiors of the outer planets.  

\begin{figure}[h]
    \centering
    \includegraphics[width=0.67\paperwidth]{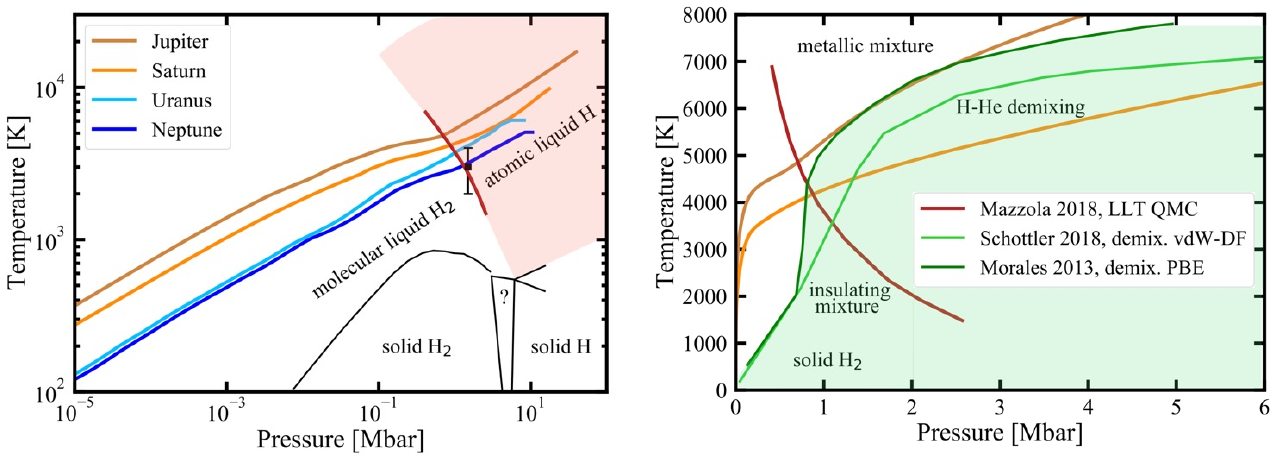}
    \caption{{\bf Left:} The phase diagram of hydrogen. Also plotted are the theoretical prediction of the fluid metallization (red line and black errorbar) in the presence of He atoms from \citet{Mazzola2018} (using electronic Quantum Monte Carlo (QMC) with the classical nuclei approximation). For reference, also shown are typical {\it P-T} profiles for Jupiter (brown), Saturn (orange), Uranus (light blue) and Neptune (dark blue) assuming adiabatic interiors. Note that although Uranus and Neptune profiles cross this molecular-to-metallic transition, it is unclear whether H-He exist in the planetary deep interiors (see Sec~\ref{sec:ice_giants} for discussion). {\bf Right:} Phase diagram for a H-He mixture predicted by numerical calculations for a mixture of proto-solar composition. Light green curve:  demixing predicted by \citet{Schoettler2018}, dark green curve: H-He demixing as predicted by \citet{Morales2013}. The shaded area represents the H-He demixing region.  The red line from \citet{Mazzola2018} is separating the insulating and metallic fluid mixture. The figures are modified from \citep{2020NatRP...2..562H}.}
    \label{fig:phasediags}
\end{figure}

\noindent{\bf Hydrogen (H): }
The EoS of hydrogen is of great importance for the gas giant planets. Under relatively low pressures, hydrogen exists as a diatomic molecular gas (H$_2$). As pressure increases, these molecules are forced closer together, leading to a transition to a liquid state. At pressures above $\sim$ 1 Mbar, and at high temperatures, hydrogen transitions from an insulating molecular liquid to a metallic liquid state \citep[e.g.,][and references therein]{Mazzola2018}. 
In this metallic state, hydrogen atoms dissociate, and electrons move freely,  resulting in a high electrical conductivity.  This transition to metallic hydrogen is relevant for giant planet interiors since the existence of metallic hydrogen leads to magnetic field generation in Jupiter and Saturn \citep[see e.g.,][and references therein]{2020NatRP...2..562H}. The left panel of Figure~\ref{fig:phasediags} shows the phase diagram of hydrogen. 

\noindent{\bf Hydrogen-Helium (H-He):}
In the deep interior of giant planets hydrogen is expected to be mixed with helium. As a result, many equations of state calculations and experiments correspond to a mixture of hydrogen and helium  in a proto-solar composition. A complexity occurs because at some pressures and temperatures helium becomes immiscible with hydrogen, leading to phase separation   \citep[e.g.,][]{stevenson1977a}. This is important for giant planet modeling because as helium becomes immiscible with metallic hydrogen, helium droplets form and settle to the planetary deep interior. This process, which is also known as "helium rain", leads to a redistribution of helium in giant planet interiors and also provides an extra source of thermal energy which affects the planetary cooling. 
The right panel of Figure~\ref{fig:phasediags} shows the predicted demixing curve of a H-He mixture \citep[for details see][and references therein]{2020NatRP...2..562H}. 

\noindent{\bf Water \& other ices:}
It is common to assume that  the giant planets consist of water (H$_2$O)  as well as other volatile materials such as ammonia and methane. Typically, "ices" are represented by water (note that the physical state of the material can be gaseous, liquid or solid depending on the exact conditions). The exact composition of the icy material is particularly important for interior models of Uranus and Neptune. 
Typically, interior models of the ice giants use EoSs of water and, when possible, of other ices such as methane and ammonia.   

\noindent{\bf Rocks and metals}:
The giant planets are also expected to consist of refractory materials such as rocks and metals. The ratio between "ices" (water, ammonia, methane) and "rocks" (silicates and metals) in the giant planet remains unknown. This value is in particular important for Uranus and Neptune since their bulk composition is dominated by heavy elements, unlike Jupiter and Saturn. To include refractory materials in planetary models often the EoSs of iron, SiO$_2$ and MgSiO$_3$ are used. 
\par 

To derive the properties (e.g., density, entropy) of a mixture, the linear mixing approximation (LMA) is used.
In Eq.~\ref{linear_mixing}, the LMA for a mixture of N elements gives for any extensive physical quantity $Q$: 
\begin{equation}
    Q(P, T) = \sum_{i=1}^{N}X_i Q_i(P, T),
    \label{linear_mixing}
\end{equation}
 where $Q_i$ is the physical quantity of the pure component $i$ and $X_i$ its mass fraction. 
For an arbitrary mixture of hydrogen, helium and heavy elements, the approximation can be written, for density and entropy, as: 
\begin{equation}
    \frac{1}{\rho(P, T)} = \frac{X}{\rho_{\rm H}(P, T)} + \frac{Y}{\rho_{\rm He}(P, T)} + \frac{Z}{\rho_{\rm Z}(P, T)},
    \label{linear_mixing_rho}
\end{equation}
\begin{equation}
    S(P, T) = X S_{\rm H}(P, T) + Y S_{\rm He}(P, T) + Z S_{\rm Z}(P, T),
    \label{linear_mixing_S}
\end{equation} 
where $X$, $Y$ and $Z$ are the  mass fractions of hydrogen, helium and heavy elements, respectively. $\rho(P, T)$ and $S(P,T)$ are the density (inverse specific volume) and specific entropy of the mixture, at a certain pressure and temperature. $\rho_{\rm H}$, $\rho_{\rm He}$ and $\rho_{\rm Z}$ are the densities of pure hydrogen, helium and heavy elements respectively while $S_{\rm H}$, $S_{\rm He}$ and $S_{\rm Z}$ are similar notations but for specific entropy. 
This implies that the range of the mass fractions is between 0 and 1 and that the sum over all mass fractions should be equal to one: $X + Y + Z = 1$. The LMA, although very useful, is not always appropriate and interactions between different species can be non-negligible \citep[e.g.,][]{hg23}. 

In addition, under some conditions different materials can undergo demixing. In that case, the materials would separate and would not be mixed together.  
This is important because it provides a physical/chemical constraint for interior models. This information can be used to identify under what conditions a modeler should assume distinct layers consisting of a single component vs.~a mixture. 


\subsubsection{Empirical interior models}
\label{sec:empirical}
A different approach for modeling planetary interiors is to use empirical models where the density profile is represented by a series of random steps in density or a mathematical function (e.g., polynomials, polytropes), and all the density profiles that match the observed properties are inferred. Rather than assuming the planet’s composition and structure and then solving for the pressure and density structure using physical EoSs, these models parameterize the structure directly and make inferences about the composition from the resulting density profiles that is set to match the observational constraints. Empirical structure models take a more unbiased view on the planetary internal structure and a priori they do not depend on knowledge of the behavior of elements at high pressures and temperatures. The advantage of such models is that they can probe solutions that are missed by the standard models, in particular, solutions that represent more complex interiors (e.g., composition gradients) with various temperature profiles (e.g., sub- and super- adiabatic). 
This approach goes back decades but has been recently revived to model the solar system giant planets. Below, when we discuss the current-state knowledge of the planetary interiors, we also discuss results from recent empirical models. 


\subsection{The interior-atmosphere connection}
\label{sec:int_atm}
As mentioned earlier, interior modelers often set the planetary outer boundary ("surface") at 1 bar which is the pressure at the surface of the Earth. Above  this pressure level the mass is negligible and will not significantly affect the calculated gravitational moments, and therefore the inferred internal structure.  On the other hand, measurements of atmospheric abundances or atmospheric phenomena such as vortices, storms, and jets, can refer to levels deeper than the 1 bar pressure level. 

Since giant planets have no surfaces, there is no clear distinction between the atmosphere and the deeper interior. As a result, the "atmospheres" of giant planets typically correspond to the region which is affected by winds and atmosphere dynamics. Also, the fact that there is no well-defined boundary between the atmospheres and the interiors, the atmosphere can be used  to reveal information on processes occurring in the planetary deep interior (heat transport, mixing, etc). 
In addition, the atmosphere is of importance since it is the radiative atmosphere which governs the cooling and contraction of giant planets. Finally, the level of enrichment provides important information on the origin and evolution histories of the planets. 
As a result, atmospheric measurements play a key role in modeling the interiors and thermal evolution of giant planets. However, despite the interplay between the atmosphere and the interior in gaseous planets, it has to be kept in mind that  atmospheric measurements may not represent the planetary bulk composition (see discussion below). 

\begin{figure}[h]
    \centering
    \includegraphics[width=0.65\paperwidth]{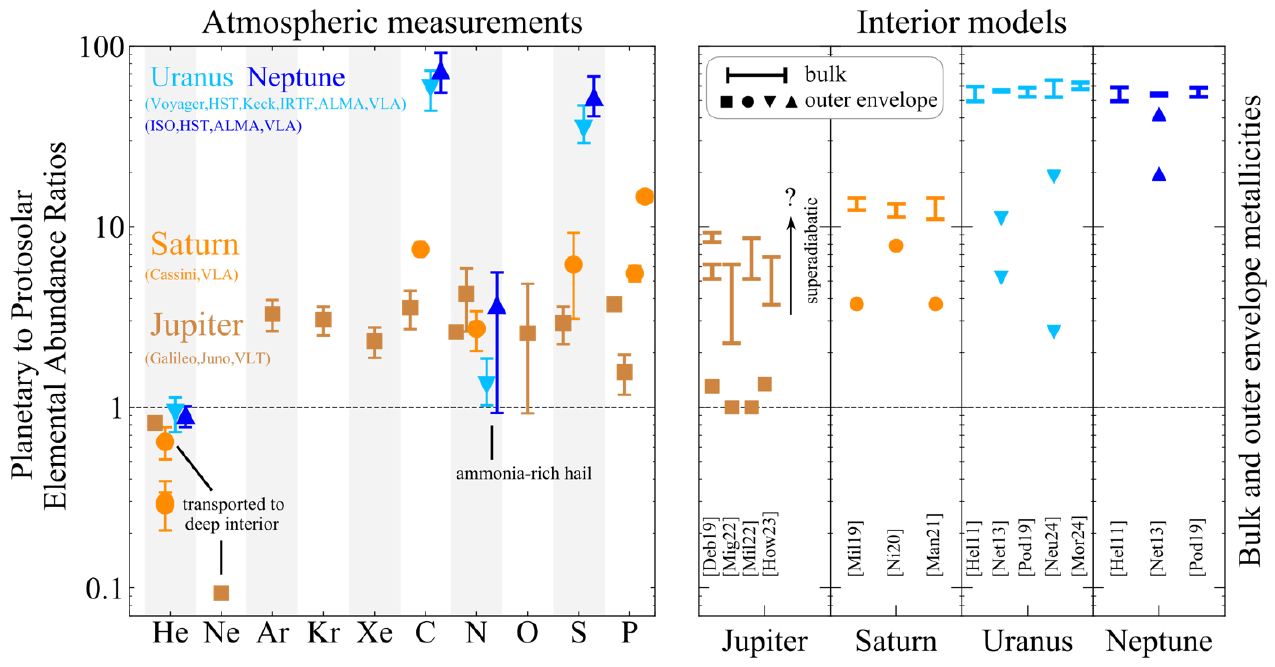}
    \caption{\textbf{Left:} Elemental abundances of He, Ne, C, N, O, P, S, Ar, Kr and Xe in proto-solar units in the atmospheres of Jupiter, Saturn,
Uranus and Neptune \citep[see][and references therein]{guillot2023}. \textbf{Right:} Bulk and outer envelope metallicities inferred from interior models, given in proto-solar units. The interior models for Jupiter are taken from  \citet{Debras2019,Miguel2022,Militzer2022,howard2023_interior}. For Saturn we use models from  \citet{militzer2019,ni2020,mankovich2021}. The solutions for Uranus and Neptune are taken from  \citet{helled11,nettel13,Podolak19,Neuenschwander2024}.
}
\label{fig:elemental_abundances}
\end{figure}

Figure~\ref{fig:elemental_abundances} presents the abundances of measured chemical elements in the atmospheres of Jupiter, Saturn, Uranus and Neptune.  These include noble gases (He, Ne, Ar, Kr, Xe), elements carried by condensible species ($\rm H_2 O$, $\rm CH_4$, $\rm NH_3$, $\rm H_2 S$), and disequilibrium species (e.g., phosphine). 
Several key conclusions can be made from the figure. First, it is clear that the atmospheres  are enriched compared to proto-solar values (left panel).  
The atmospheric enrichment is different for the four planets: it is a factor of $\sim$3 for Jupiter, about 7 for Saturn, and between 30 and 90 for Uranus and Neptune. As we discuss  in more detail below, the bulk metallicity of a giant planet is determined by its formation and evolution history. 
Currently, it is still poorly understood whether (and under what conditions) the measured atmospheric enrichment represents the planetary bulk composition \citep[see e.g.,][and dicussion below]{HelledStevenson2022}. The right panel shows the planetary bulk metallicities derived   from interior models (discussed in Sec.~\ref{chap1:sec2}). For the four giant planets, such models infer an outer envelope metallicity lower than predicted by atmospheric measurements. This suggests that the outer envelope metallicity is not representative of the bulk metallicity. 
Second, it is clear that the level of available information is very different for the different planets: while we have detailed information for Jupiter thanks to the Galileo  probe, and some information for Saturn, we have only a few elements measured for Uranus and Neptune with rather large uncertainties. This highlights the importance of sending atmospheric probes to measure the enrichment of various elements down to several to tens of bars in the  atmospheres of giant planets.   
Third, helium is slightly depleted in Jupiter's atmosphere. The precise helium content of Saturn's atmosphere is unknown  but is expected to be even more depleted \citep{Mankovich2020}. This is not because helium is overall depleted in the gas giant planets compared to proto-solar composition, but instead a hint for the process of helium rain (see Sec.~\ref{sec:eos}). 
\par


\subsection{The composition-formation  connection}
\label{sec:formation}
In the standard model for giant planet formation, also known as core accretion,  a giant planet is formed via three main phases. 
The earliest stage of giant planet growth is core formation (phase 1), which includes the accretion of  solids (e.g., pebbles, planetesimals) to build up a heavy-element core. During this early phase some H-He gas from the protoplanetary disk can be accreted but the surrounding gas envelope remains small in mass. Once the core mass reaches about $\sim 2 M_{\oplus}$ the accreted solids can vaporize in the planetary envelope. 
Phase-2 is characterized by a slower heavy-element accretion and a steadily increasing gas accretion rate. Composition gradients can form during this phase, especially as the rate of H-He  accretion becomes comparable to that of the heavy elements, marking the transition to runaway gas accretion (phase-3).  
During phase-3, gas accretion occurs so rapidly that the disk cannot supply gas fast enough to keep up with the planet's contraction.  During this final stage of gas accretion, a giant planet acquires most of its mass and becomes dominated by H-He in composition.   In the case of Uranus and Neptune (and possibly also Saturn), the planetary formation ends in phase-2 and phase-3 of runaway gas accretion does not take place. 
\par

\begin{figure}[h]
    \centering
\includegraphics[width=0.45\paperwidth]{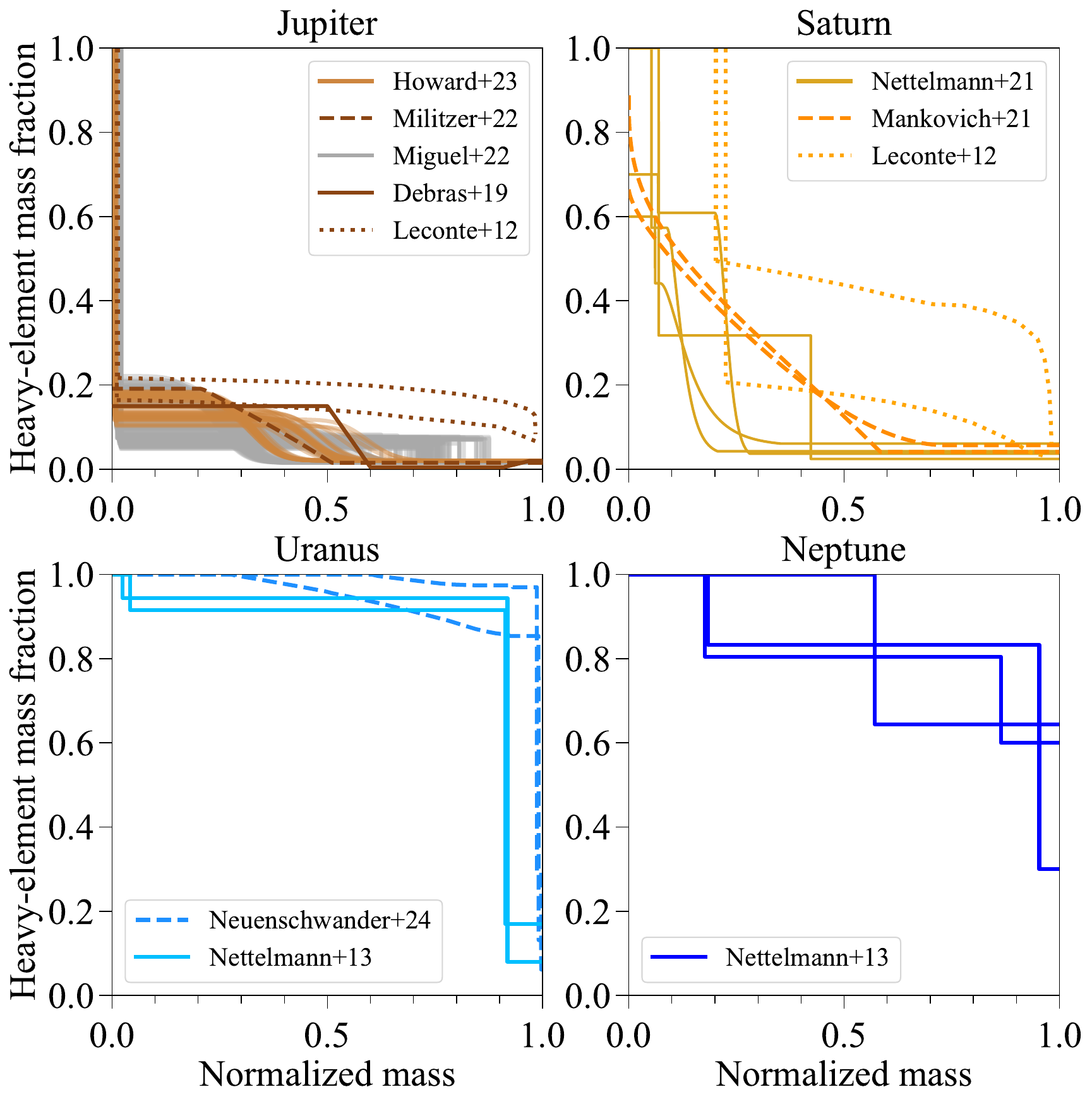}
    \caption{The distribution of heavy elements vs.~normalozed mass in Jupiter, Saturn, Uranus and Neptune from various published models. Howard+23: \citet{howard2023_interior}. Militzer+22: \citet{Militzer2022}. Miguel+22: \citet{Miguel2022}. Debras+19: \citet{Debras2019}. Leconte+12: \citet{Leconte2012}. Nettelmann+21: \citet{nettelmann2021}. Mankovich+21: \citet{mankovich2021}. Neuenschwander+24: \citet{Neuenschwander2024}. Nettelmann+13: \citet{nettel13}. Multiple models can be displayed per reference when various distributions of heavy elements are inferred.}
    \label{fig:Z}
\end{figure}

The paradigm of a mixed interiors in Jupiter and Saturn (due to convective mixing) has recently been supplanted by the idea of the presence of compositional gradients in their envelope \citep[e.g.,][and Sec.~\ref{sec:gas_giants}]{Wahl2017,Miguel2022,mankovich2021}. Such compositional gradients may also exist in the interiors of Uranus and Neptune \citep[e.g.,][and Sec.~\ref{sec:ice_giants}]{Vazan2020}.  
Formation models do predict that giant planets are enriched in heavy elements compared to proto-solar composition  but at the same time they also predict that the planets are inhomogeneous with the deep interior being more enriched with heavy-elements in comparison to the atmosphere, a prediction that is consistent with interor models. In a way, the distribution of the heavy-elements in the planetary interior reflects the ratio between the heavy-element and gas accretion rates during the planetary formation. As a result, the metallicity of the atmosphere is not expected to represent the planetary bulk metallicity, unless the planet is fully convective. Figure~\ref{fig:Z} shows the distribution of heavy elements in the four planets inferred from interior models. They suggest the presence of compositional gradients and confirm that the atmospheric metallicity is not representative of the bulk metallicity as the deep interior appears to be more enriched. 
\par 

In addition, in the case of Jupiter, formation models find that the accreted gaseous envelope is unlikely to be metal-rich unless there is a subsequent phase where a small amount of solids (a few M$_{\oplus}$) is accreted to account for the observed atmospheric enrichment. However, this remains under investigation  \citep[see][and references therein for discussion]{Helled2022b,HelledAGU2024}. This raises the question of what mechanisms can lead to the enrichment measured in Jupiter's atmosphere. One potential explanation is the existence of a stable region separating the atmosphere from the envelope, which could lead to a partitioning of the heavy-element content \citep{howard2023_invZ,2024ApJ...967....7M}. The connection between the atmospheric and bulk composition of the other outer planets and giant exoplanets is still being explored.  
Overall, the realization that the heavy-element is expected to change with depth in the interiors of giant planets (i.e., the interior is non-homogeneous in composition) adds a complexity to understanding the connections between atmospheric composition, bulk composition, and the planet's origin and internal structure. 

\subsection{The evolution of the planetary  interior}
\label{sec:evolution}
Giant planets emerge from their formation as hot, luminous, and extended objects. Subsequently, they cool down and contract over billions of years. This  planetary long-term thermal evolution  is key to connect planet formation and the present-day structure. In this context, evolutionary models are necessary to further constrain formation and interior models.

Often, evolution models aim to match the planet's  effective temperature $T_{\rm eff}$ at the present day. $T_{\rm eff}$ can be linked to the planetary luminosity from the following relation:
\begin{equation}
    L = 4 \pi \sigma R^2 T_{\rm eff}^4,
\end{equation}
where $\sigma$ is the Stefan-Boltzmann constant and $R$ the planetary radius. Although the luminosities of the solar system giant planets can be measured, it is challenging to accurately determine the fraction originating from a planet's deep interior (intrinsic luminosity) versus the fraction reflected from stellar irradiation. 
The energy balance equation is given by:
\begin{equation}
    L = L_{\odot} + L_{\rm int},
\end{equation}
where $L_{\odot}$ is the re-radiated absorbed solar power and $L_{\rm int}$ is the intrinsic luminosity which corresponds to the planet internal heat loss. A key ingredient in the energy balance equation is the Bond albedo, which is the measure of the total fraction of incident electromagnetic radiation that the planet reflects back into space over all wavelengths and angles (see Table~\ref{tab:properties_planets}). 
In order to model the evolution of the planetary interior with time, we add the time-dependent equation to the structure equations showed in Sec.~\ref{sec:make_model}:
\begin{equation}
    \frac{dL}{dr}=4 \pi r^2 \rho \left( \dot{\epsilon} - T \frac{dS}{dt}\right),
    \label{eq:luminosity}
\end{equation}
where $t$ is the time and $\dot{\epsilon}$ is a source term (e.g., nuclear reactions, radioactivity, or energy acquired by accretion) often assumed to be 0 for planets. Starting from an initial state, we can integrate Eq.~\ref{eq:luminosity} and therefore follow the evolution of the parameters of the planet.

\begin{figure}[h]
    \centering
    \includegraphics[width=0.65\paperwidth]{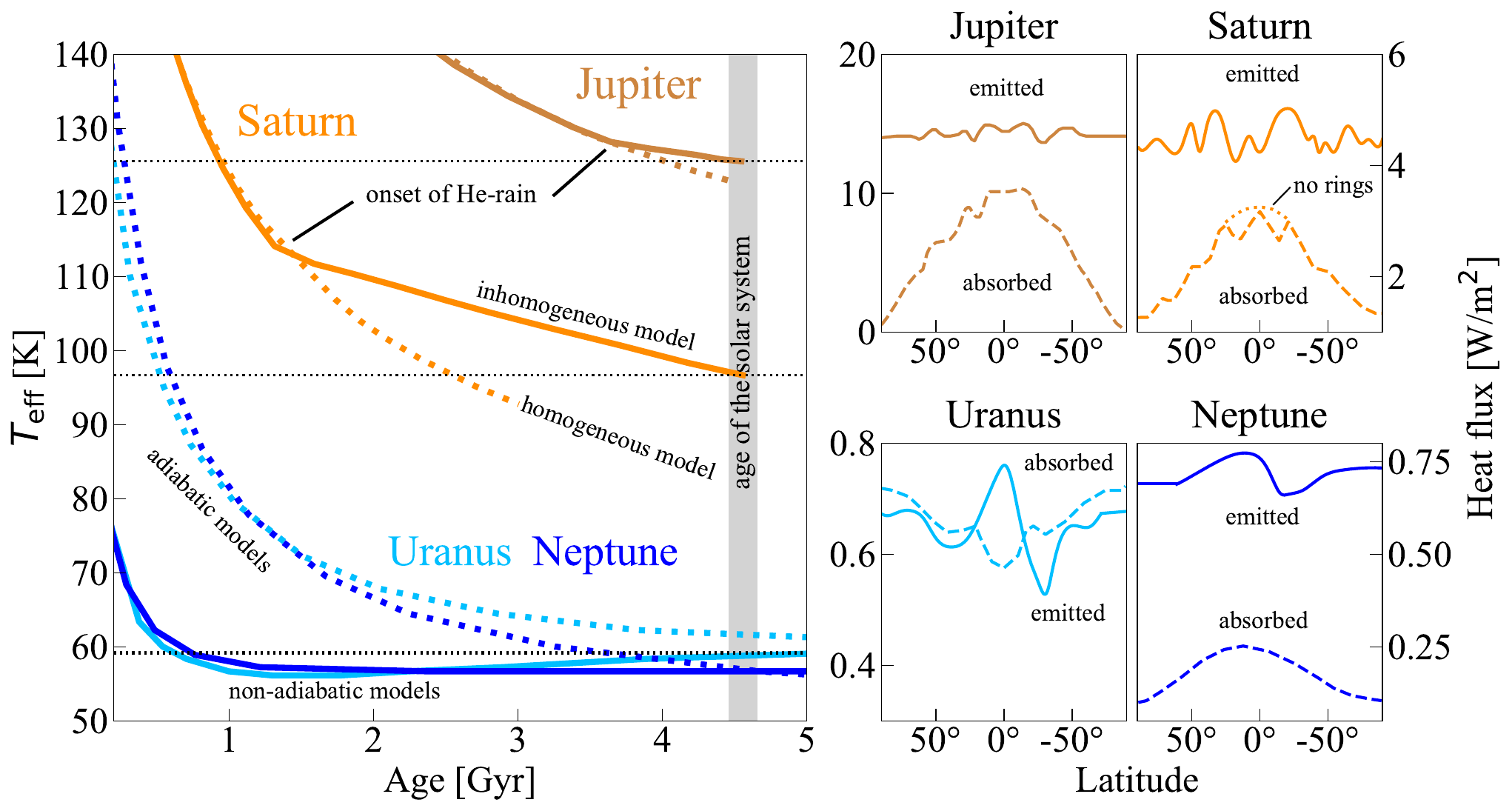}
    \caption{{\bf Left:} The thermal evolution of the giant planets. Jupiter and Saturn models were taken from \citet{Mankovich2020}; Uranus and Neptune models were taken from \citet{Scheibe2021}. For Jupiter and Saturn, dotted lines show homogeneous evolutions while solid lines show evolutions with demixing of helium. Helium rain starts at $\sim$3.5~Gyr for Jupiter, and at 1.3~Gyr for Saturn.  For Uranus and Neptune, dotted lines show adiabatic models while solid lines show non-adiabatic models. {\bf Right:} Heat fluxes of the four giant planets as a function of latitude (adapted from \citep{beatty1999new}). The solid curves show the emitted flux while the dashed ones show the absorbed flux (see text for discussion).} 
    \label{fig:evolutions}
\end{figure}

The left panel of Figure~\ref{fig:evolutions} shows results for evolution calculations of Jupiter, Saturn, Uranus and Neptune. The dotted and solid lines correspond to homogeneous (i.e. constant composition over time) and non-homogeneous evolution, respectively. 
For Jupiter, we can see from the figure that the assumption of a homogeneous evolution works fairly well \citep[e.g.,][]{hubbard1969}, although, as we discuss below, Jupiter's interior is non-homogeneous. The figure clearly shows that Saturn's current luminosity is higher than predicted by homogeneous evolution models.   
Differentiation of helium, due to helium rain, can provide an additional source of energy to explain Saturn's excess of luminosity \citep[e.g.,][]{stevenson1977a}.
Actually, the measurement of a slight depletion of helium by the Galileo probe confirmed that this phenomenon also occurs in Jupiter. Helium differentiation is hence expected to be more pronounced in Saturn in order to yield the correct cooling time, as shown by the solid orange line in the left panel of Fig.~\ref{fig:evolutions}. 
For Uranus, adiabatic models lead to a higher $T_{\rm eff}$ compared to the measured one, while for Neptune,  an adiabatic evolution can explain its measured effective temperature. Non-adiabatic models of both planets, however, could be consistent with the observed values (see section \ref{sec:ice_giants} for further discussion).  
\par

The right panel of Figure~\ref{fig:evolutions} shows the energy balance of the four planets by comparing the absorbed and emitted
heat fluxes. First, it is clear that Uranus is the only planet that is in thermal equilibrium with the sun, i.e., the emitted flux is of the same order as the absorbed flux. The other three planets, on the other hand, emit more energy than they receive from the sun. This could imply that they have an internal energy source: in the case of the giant planet, this is not an energy source like nuclear reaction, but remaining primordial energy and/or helium rain. It could also be a result of inefficient cooling that traps the heat in the planetary deep interior. Traditionally, the higher heat fluxes of Jupiter, Saturn and Neptune were interpreted as being the signatures of convective interiors since heat is being efficiently transported from deep hot regions. 
Second, it is clear that most of the flux is absorbed at the equatorial region (besides Uranus due its its unusual tilt) but is emitted relatively uniformly across latitude, suggesting efficient heat transport (i.e., convective e mixing) in the atmospheres.  Finally, it is noticeable that the emitted fluxes of Jupiter and Saturn are significantly higher than that of Uranus and Neptune.


\section{Current-state knowledge of the outer planets}
\label{chap1:sec2}

\begin{figure}[h!]
    \centering
\includegraphics[width=0.75\paperwidth]{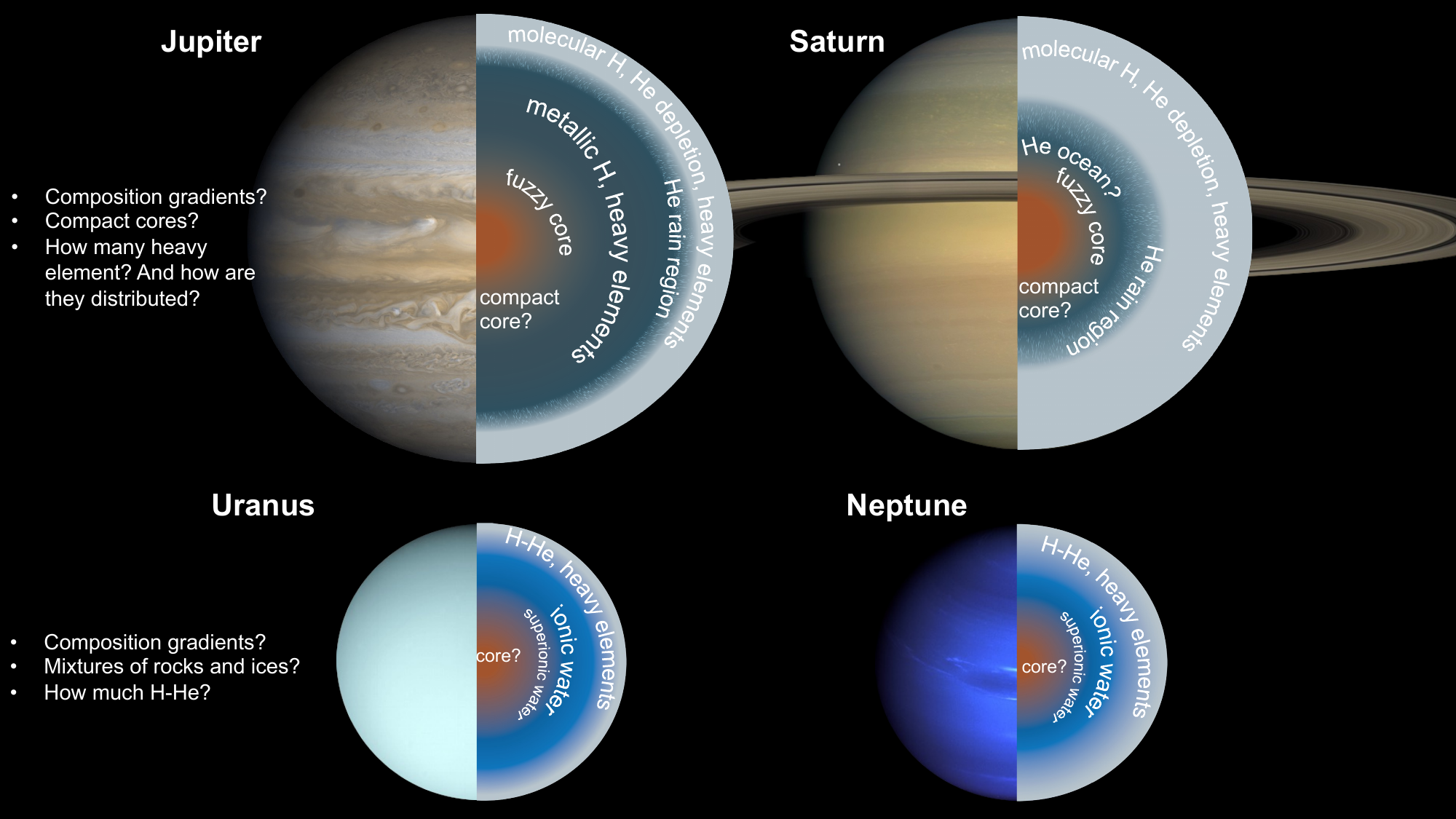}
    \caption{Sketches of the internal structures of Jupiter, Saturn, Uranus and Neptune.}
    \label{fig:sketches}
\end{figure}

Section~\ref{chap1:sec1} presented how to make an interior model, relying either on EoSs or on an empirical approach. It already showed, although a bit prematurely, some results of the interior structures of the outer planets inferred from structure models. The bulk metallicities and distributions of heavy elements inferred from interior models are shown on Figures~\ref{fig:elemental_abundances} and~\ref{fig:Z} and the planetary evolution are shown in Figure~\ref{fig:evolutions}. In this section, we summarize the current-state knowledge of the four outer planets. Figure~\ref{fig:sketches} sketches their internal structures and outlines several key questions that are currently being  investigated.  The results of interior and evolution models presented in this section  explain how we arrived at this understanding. 

\begin{figure}[h]
    \centering
\includegraphics[width=0.45\paperwidth]{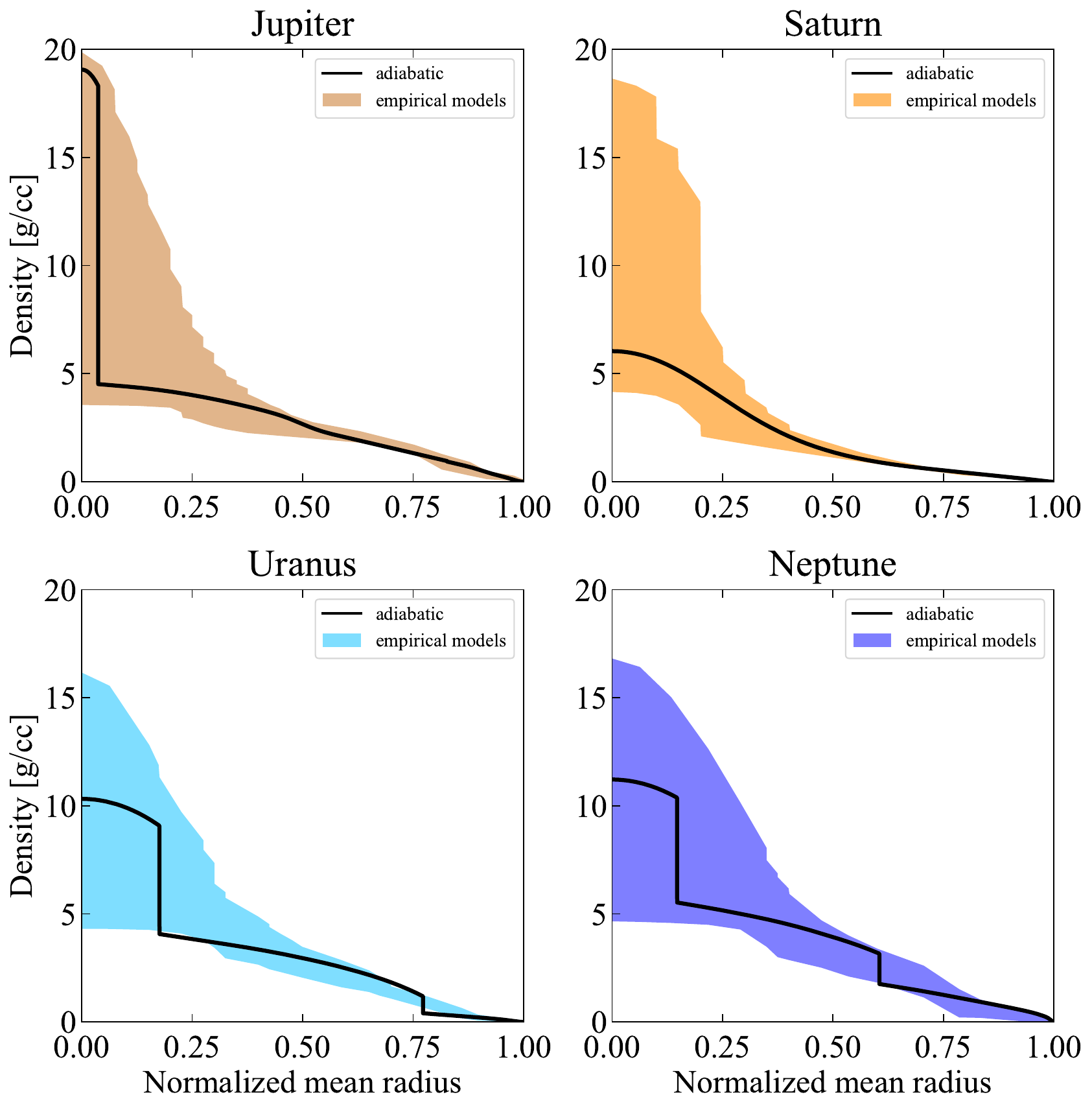}
    \caption{Density profiles of Jupiter, Saturn, Uranus and Neptune. The shaded areas show the range of solutions from a set of empirical models, taken from \citet{Neuenschwander2021} for Jupiter, from \citet{2020ApJ...891..109M} for Saturn and from \citet{Neuenschwander2022} for Uranus and Neptune. For comparison, for each planet we also show one adiabatic model, taken from \cite{howard2023_interior} for Jupiter, from \citep{mankovich2021} for Saturn, and from \citep{nettel13} for Uranus and Neptune (models U1 and N2b).}
    \label{fig:rho}
\end{figure}

\newpage
Modeling the planetary interior is non-unique since many  solutions can lead to the same observed properties and the parameter space of valid structures is rather wide. 
Today planetary modelers can use advanced  statistical methods to explore the large numbers of possible solutions for the planetary interiors such as Markov chain Monte Carlo (MCMC) algorithms based on Bayesian inference.  
Figure~\ref{fig:rho} shows a large range of density profiles for Jupiter, Saturn, Uranus and Neptune as inferred from empirical models obtained by MCMC sampling. For comparison, we also present for each planet a solution based on canonical interior models. 

\subsection{The "gas" giant planets}
\label{sec:gas_giants}
Jupiter and Saturn are known as the gas giants in the solar system. This is because their large masses of 317.83 and 95.16 $M_{\oplus}$,  respectively, making them "giants" and their inferred composition which is dominated by hydrogen and helium, making them "gas giants".
Note that the name "gas giants" is somehow misleading, since most of the planetary interior is "fluid" and not "gas" \citep[e.g.,][and references therein]{2020NatRP...2..562H}. Although Saturn's mass is only about one-third that of Jupiter, its mean radius is 9.14 times that of Earth, compared to Jupiter's mean radius of 10.97 times Earth's radius. This relatively small difference in radii, despite the significant difference in mass, can be attributed to Saturn's higher enrichment of heavy elements. In addition, by comparing the moment of inertia (MoI) values of the two planets, one can conclude that Saturn is more centrally condensed as its MoI value is smaller than that of Jupiter (see discussion below).
\par 

Despite the accurate measurements of their gravitational fields by the Juno and Cassini missions, the exact distribution and total mass of heavy elements in Jupiter and Saturn remain unknown. 
Recent models of Jupiter and Saturn's structures reveal complex interiors characterized by composition variations, non-convective zones, and fuzzy cores \citep{HelledStevenson2022,HelledAGU2024}.  
Furthermore, it is now clear that significant distinctions exist between Jupiter and Saturn, emphasizing the uniqueness of each  planet. 

\subsubsection{Jupiter}
\label{sec:jupiter}
Jupiter's internal structure has been studied for decades. Actually, this year, 2024, marks the 100th anniversary of the first published model of Jupiter's interior by \citet{jeffreys1924}. At that time, the presence of hydrogen and helium inside the largest planet in our solar system was already under consideration. Since then, great progress in theory, in particular on our understanding of hydrogen at high-pressure, coupled with accurate data, notably on gravity field measurements, improved our comprehension of Jupiter's interior.  Surprisingly, Jupiter's internal structure remains illusive. This is, however, in part due to this great progress which introduced new challenges and provided additional data that have to be combined to provide a consistent and comprehensive view of the planet. Current Jupiter models include a non-homogeneous interior, composition gradients, and account for the dynamical contribution when calculating the density profile. 
\par 

Traditional interior models of Jupiter generally consisted of a central compact core (purely made of heavy elements, i.e. ices and rocks), surrounded by an envelope of hydrogen and helium. 
The core was assumed to exist based on giant planet formation models at the time that assumed that gas accretion is initiated once the core reaches masses of a few 10 M$_{\oplus}$.  
As the pressure increases, hydrogen transitions from molecular to metallic state (metallic hydrogen). Initially, the planetary envelope was thought to be homogeneous \citep{hubbard1969} and composed of a  proto-solar mixture. 
Then, differentiation of helium within the interior was suggested   as helium was found to be immiscible with metallic hydrogen \citep{stevenson1977a}. Hence, the so-called 3-layer picture  emerged: a heavy-element core, surrounded by a He-rich layer of metallic hydrogen and a He-poor layer of molecular hydrogen. 
\par

Thanks to the high accuracy of Juno's measurements more complex models of Jupiter have been developed. These models account for the winds and include the dynamical contribution to the even gravitational moments. 
Recent Jupiter models that fit Juno clearly show that Jupiter's interior has a non-uniform distribution of heavy elements, with a greater metallicity in the deeper regions \citep{Wahl2017,Debras2019,nettelmann2021,Miguel2022, howard2023_interior}. Furthermore, Jupiter's core is expected to be "fuzzy", suggesting that the central region is not composed of only heavy elements, but also include lighter elements like H and He. In addition, the core is not compact but is extended and can be up to tens of percents of  Jupiter's radius. 
Finally, these updated models also indicate significant deviations from the traditional  adiabatic temperature profile in the deep interior. Depending on the model, Jupiter's inferred total heavy-element mass varies between 18 and 45 M$_{\oplus}$ \citep{Debras2019,ni2019,Miguel2022,Militzer2022,howard2023_interior}. Note that the existence of heavy-element gradients in Jupiter's deep interior may lead to higher metallicities as it can lead to a super-adiabatic temperature profile. While adiabatic models typically infer a central temperature of $\sim 20,000~$K, superadiabatic models yielded values of $\sim 35,000~$K \citep{Debras2019,Militzer2022}. It should also be kept in mind that all these interior models rely on specific assumptions, and in some cases modifications to the 1-bar temperature, modifications to the EoS and optimized wind models must be used to fit the available data. 

Currently, is still unknown whether Jupiter possesses a small pure heavy-element core of a few M$_{\oplus}$.  Updated giant formation models suggest that the pure heavy element core can be very small, up to a few M$_{\oplus}$ \citep[see][and references therein]{HelledAGU2024}.  
In addition, most of the interior models of Jupiter infer a rather low metallicity in the outer envelope (solar or slightly supersolar). As discussed before, this is  inconsistent with the atmospheric enrichment measured as shown in Fig.~\ref{fig:elemental_abundances}. 
\par 

\noindent \underline{\bf Results from empirical models:}\\
\citet{Neuenschwander2021} presented a large range of empirical models of Jupiter where the density profile is  represented by (up to) three polytropes.  They 
 investigated the relationship between Jupiter's moment of inertia (MoI) and the gravitational moments $J_2$ and $J_4$. It was shown that a measurement of Jupiter's MoI, with an accuracy better than 0.1\%, can further constrain Jupiter's internal structure. It was also  found that models with a density discontinuity at $\sim$ 1 Mbar, as predicted by H-He phase
separation simulations, have low central densities, corresponding to a fuzzy core in Jupiter. 

\subsubsection{Saturn}
\label{sec:saturn}
Modeling Saturn's interior presents challenges, but different from those associated with modeling Jupiter. 
Also in the case of Saturn, significant improvements in interior models became possible thanks to the Cassini mission which provided accurate gravity \citep{Iess2019} but also ring seismology data \citep[e.g.,][]{Fuller2014}. Regarding the models' structure and its different layers, similar assumptions used to be made as those applied to Jupiter. We present here the main and latest results of Saturn's models and discuss how it may differ from Jupiter. 
\par

\citet{militzer2019} modeled Saturn's internal structure using Monte Carlo sampling and found a relatively massive pure heavy-element core with a mass  between 15 and 18 M$_{\oplus}$, with a moderate envelope enrichment corresponding to $<$ 5 M$_{\oplus}$ of heavy elements. 
Assuming a 4-layer model for Saturn yielded a total heavy-element mass of 12–18 M$_{\oplus}$, assuming a core surrounded by a fuzzy region \citep{ni2020}. \citet{mankovich2021} presented interior models of Saturn that fit both the  gravitational moments and oscillation modes from ring seismology. Saturn's interior was found to possess a large region that is stable  against convection which extends to $\sim$60\% of Saturn's radius. The core region was found to have a moderately low central density ($\sim$6 g/cm$^3$), indicative of a fuzzy core. \citet{nettelmann2021}  found a total amount of heavy elements of $12.6-13.6~M_{\oplus}$ (compared to $7.5-10.1~M_{\oplus}$ for Jupiter).
When changing the  H-He EoS, an enriched envelope extending to $<$ 0.4 of Saturn's radius and compact heavy-element cores can also be inferred. When considering EoS perturbations, Saturn was found to have a fuzzy core extending to 40\% of Saturn's radius, a result which is consistent with the models of \citet{mankovich2021}. This structure model suggests a dilute core and helium rain in Saturn, indicating that it has an inhomogeneous interior and regions that are stable against convection. The presence of a fuzzy core is hence suggested in both Jupiter and Saturn. However, their structure differs: Saturn's fuzzy core is about 60~$M_\oplus$ and the mass fraction of heavy elements inside can reach 80\% while Jupiter's fuzzy core is of the order of $\sim 100~M_\oplus$ and is still dominated by H-He \citep{HelledAGU2024}. 
\par

Thermal evolution models can also serve as valuable complements to interior models.  As discussed in Sec.1.3., while helium rain probably started only relatively recently in Jupiter, it is expected to play a more important role in Saturn's interior and evolution. Indeed, as Saturn is cooler, state-of-the-art H-He phase diagrams \citep[e.g.,][]{Morales2013,Schoettler2018} predict helium demixing to occur much earlier. This phenomenon is expected to be significant in Saturn and affects a larger fraction of the planet's interior. Hence, recent evolution models of Saturn \citep[e.g.,][]{Mankovich2020} suggest a stronger depletion in helium of the atmosphere, compared to Jupiter. The helium gradient established in the present-day interior of Saturn could range  from a strong helium depletion ($Y<0.1$) in the outer envelope to a very He-rich mixture ($Y \sim 1$) in Saturn's deep interior. Substantial helium rain could lead to the formation of a helium ocean (or helium shell) above the central heavy-element core of Saturn. It should be noted, however, that such evolution calculations strongly depend on the H-He phase diagram which remain very uncertain. 

\noindent \underline{\bf Results from empirical models:}\\
For Saturn, \citep{2020ApJ...891..109M} applied a Bayesian
MCMC-driven approach to explore the full range of possible density distributions
in Saturn where the density profiles were represented with piecewise polynomials. These models exhibit high densities in the outer parts of the planet, suggesting significant heavy element enrichment (as expected), and while the inner half of Saturn was less well constrained most models exhibit significant density enhancement (interpreted as a core) but with density values consistent with a dilute, rather than a compact heavy-element core. 


\subsection{The "ice giants"}
\label{sec:ice_giants}
Uranus and Neptune, the so called "ice giants", have masses of 14.5 and 17 M$_{\oplus}$, respectively, and  sizes of about 3.98 and 3.86 times Earth's mean radius. These planets are unique as they are significantly smaller than the gas giants, but significantly larger than terrestrial planets. Also, in terms of composition, they differ fundamentally from the other two groups, since their composition is expected to be dominated with heavy elements, but probably including volatile materials, and also consist of H-He atmospheres. However, unlike the gas giants  Jupiter and Saturn, H-He is expected to represent only about 10-15\% of the total planetary mass in Uranus and Neptune \citep[see ][and references  therein]{HelledFortney2020}.  
 As Uranus and Neptune have only been explored by the Voyager 2 flybys in 1986 and 1989, questions regarding the evolution, the atmospheric composition, or the internal structure and many more remain unanswered.  
 \par 
 

The bulk compositions and internal architectures of Uranus and Neptune remain elusive although their mean densities (1.27 g cm$^{-3}$ for Uranus and 1.64 g cm$^{-3}$ for Neptune), suggest that they consist of heavier elements than H-He. These substances could be water, ammonia, and methane, which are referred to as "ices" but they could also have significant amounts of rocks. Despite some apparent similarities, such as mass and radius, Uranus and Neptune also exhibit fundamental differences such as significantly different satellite systems, luminosties, and inferred MoI values. At this stage, it is still unclear how similar the two planets are when it comes to their internal structures. It is also unclear whether Uranus and Neptune are predominantly water-dominated ("icy") or rock-dominated ("rocky"), as the inferred global water-to-rock ratio varies widely and is heavily dependent on the model assumptions.  Thus, the name "ice giants" for Uranus and Neptune might not reflect their bulk compositions \citep{Helled2020,HelledFortney2020}. Uranus and Neptune could actually be "rock giants" instead of "ice giants," with interiors composed of substantial amounts of silicates 
\citep[see][and references therein]{Neuenschwander2022,Neuenschwander2024}. 


\subsubsection{Uranus}
\label{sec:uranus}
The bulk composition of Uranus is not well constrained but is expected to be dominated by heavy elements.
Traditional three-layer structure models of Uranus propose that they are composed of up to  $\sim$ 2 $M_{\oplus}$ of H-He \citep[e.g.,][and references  therein] {helled11,nettel13,Neuenschwander2024}. 
\cite{nettel13} modeled Uranus assuming a rocky core surrounded by two adiabatic and homogeneous envelopes of H-He and water. The outer envelope metallicity was found to be  $Z\leq17\%$ assuming the Voyager rotation period and $Z\leq8\%$  when using the modified shape and rotation rate inferred by \citep{Helled2010shape}. However, the inner envelope is found to be significantly enriched in heavy elements, with $Z \sim 0.9$.  It should be noted, however, that solutions with solar metallicity ($Z=0.015$) were also found to fit the available data. \cite{nettel13} finds that Uranus' luminosity can not be explained by assuming an adiabatic interior. In such adiabatic models, the central temperature of Uranus is about $5000-6500~$K. 

Models by \cite{Bailey2021} suggest that Uranus' envelope metallicity is $\lesssim$  0.01, potentially suggestive of fully demixed hydrogen and water. 
By assuming hydrogen-water immiscibility it was suggested that the low observed heat flow of Uranus could be a result of this process. 
It should be noted that the volatile materials in Uranus are typically represented by water although methane and ammonia are also expected to exist. Recent models by \citet{2024arXiv240312512M} suggest that the interiors of Uranus and Neptune are methane-rich.
\par 


Hints on Uranus' interior can come from evolution models. Uranus' low luminosity implies that the interior is not fully convective/adiabatic. The consideration of a non-adiabatic temperature profiles affects the inferred planetary composition and therefore must be considered when modeling Uranus' internal structure. \cite{Vazan2020} showed that a composition gradient in Uranus' interior naturally explains its low luminosity, without the need of artificial thermal boundaries. Composition gradients inhibit convective mixing and as a result can and slow down the planetary cooling. In addition, it was shown that two- and three-layer models of Uranus are able to fit Uranus properties only if the interior is very cold and (partially) conductive \cite{Vazan2020}. Since non-adiabatic interior models are hotter, they lead to a higher planetary metallicity\footnote{because if the temperature is higher more denser elements can be incorporated and result in the same density}, with Z$\sim$0.95 (where cold models typically have Z$\sim$0.95). Similarly, \cite{Scheibe2021} showed that a Uranus with thermal boundary layers can explain the  observed present-day luminosity. However, this is possible   with the existence of an interface thick enough to trap the majority of Uranus' primordial heat in the deep interior so that the outer envelope is in equilibrium with the sun's irradiation. 
In that case Uranus' deep interior remains hot and a temperature jump of the order of $\sim$ 8000 K develops between outer and inner envelope. The high central temperature of $> 2 \cdot 10^4$ K is similar to the ones inferred by \cite{Vazan2020}. 


\noindent \underline{\bf Results from empirical models:}\\
Given that  the interiors of Uranus and Neptune are poorly constrained, empirical  models offer an interesting path to investigate its internal structure. Relaxing the assumption of an adiabatic (fully convective) interior suggests that Uranus has no distinct layers, and that the planetary interior consists of composition gradients and/or boundary layers. These models also challenge the notion that Uranus' interior is water-rich \citep{helled11,Neuenschwander2024} and typically infer higher central temperatures, going up to 50,000 K. For comparison, adiabatic models typically infer cenral temperatures of the order of several 1000 K. While the \citet{nettel13} models are very water-rich, in fact, the rock-to-water ratio is unknown and solutions where Uranus' interior is rock-rich have also been presented.  For example, \citet{helled11} showed that a rock-dominated Uranus (with a mixture of rock-H-He) can have 10.9 M$_{\oplus}$ of rocks (represented by SiO$_2$), while a water-H-He interior has 12.8 M$_{\oplus}$ with the rest being H-He. 

\citep{helled11} inferred a metallicity of $\sim$76\% to $\sim$ 90\% for Uranus when the heavy elements are being represented by SiO$_2$ (rock) and H$_2$O (water), respectively. \citet{Podolak19} showed that the inferred densities profiles are consistent with compositions gradients, and therefore with non-adiabatic temperature profiles. 
\citet{Movshovitz22} used empirical models to investigate how the density profiles of Uranus and Neptune can be further constrained with improved measurements of their gravitational moments. It was also found that a fraction of Uranus’ envelope is consistent with an adiabatic region of H-He with solar atmospheric abundances \cite{Movshovitz22}.
\cite{Podolak2022} used empirical models with an algorithm to determine the temperature profile and showed that the rock-to-water ratio in Uranus can't be much larger than two. However, in this work only 2 components per layer were considered simultaneously and isothermal profiles were preferred over adiabatic profiles. 
\cite{Neuenschwander2024} used empirical models to identify  non-adiabatic regions in Uranus’ interior. This leads to significantly hotter internal temperatures that can reach a few $10^3$ K and higher bulk heavy-element abundances (by up to 1 M$_{\oplus}$) compared to standard adiabatic models. They also explored how the assumed rotation period affects the inferred composition. Finally, they showed that while  solutions with only H-He and rock are possible, the maximum rock-to-water ratio was found to range between 0.05 and 0.38. This is significantly lower compared to standard adiabatic models. Finally, the authors predicted the higher-order gravitational coefficients ($J_6$ and $J_8$) for the planets based on various assumptions regarding the planetary rotation periods, wind depths, and uncertainties in lower-order harmonics. 
Finally, \cite{2024arXiv240312512M} propose that a significant fraction of accreted planetesimals by Uranus included highly carbon-rich refractory organic materials, similar to composition of comets. The rock-to-water ratio in such models span a considerable range of values, but since a large fraction of methane is also present these models always contain more ice (methane and water) than rock and iron. Thus, the interior of the ’ice giant’ Uranus would indeed be strictly icy.


\subsubsection{Neptune}
\label{sec:neptune}
Typically, interior models of Uranus and Neptune have been developed together. While their inferred internal structure are found to be somewhat similar, there are also important differences. First, the uncertainty on the gravitational moments of Neptune is larger than Uranus and therefore the parameter space of possible solutions is broader. The measured high heat flux of Neptune might hint that heat transport in Neptune is more efficient than in Uranus. The fact that Neptune's inferred MoI value is larger than that of Uranus, suggest that of Uranus suggests that the planet is less centrally condensed, which might hint for more efficient convection. Still, Neptune could have composition gradients and boundary layers. Unfortunately, our understanding of how the planets differ from each other is still very poor. 


Three-layer adiabatic models of Neptune have an outer envelope metallicity of $\sim$ 55-65\%, depending on the assumed rotation period. Like in the case of Uranus, also here envelopes with solar metallicities are possible \cite{nettel13}. Neptune's luminosity is higher than Uranus' and is somewhat more consistent with a convective/adiabatic interior. Indeed, Neptune’s luminosity can, in principle, be reconstructed when assuming an adiabatic interior \cite{nettel13}. 
For Neptune, two possible solutions with boundary layers have been found \cite{Scheibe2021}. Neptune could either have a boundary layer with a thickness larger than 15 km. 
It was suggested that Neptune's inner envelope is cooling where the temperature gradient between the outer and inner envelope is decreasing with time (Gyrs timescale). In this  scenario, Neptune's central temperatures are of the order of 10,000-20,000 K. Another possible solution is a narrow boundary (thickness $<$15 km) where the thermal boundary layers of enhanced conductivity is around $\lambda/\lambda_{H_{2}O} \sim$ 100, perhaps suggesting a layered double-diffusive state. Energy transport is efficient and the planet can appear brighter than in the standard  adiabatic case \cite[see][for further details]{Scheibe2021}. 

\cite{Bailey2021} suggested that Neptune's envelope contains a substantial water mole fraction, as much as $\gtrsim 0.1$ relative to hydrogen. When assuming hydrogen-water immiscibility it was suggested that gravitational potential energy release due to present-day hydrogen-water demixing could explain Neptune's observed heat flow, if convection in its deep interior is inhibited similarly to the case of Uranus. 

\noindent \underline{\bf Results from empirical models:}\\
For Neptune, \citet{helled11} inferred heavy-element masses of 15.5 M$_{\oplus}$ (Z=0.904) and 13.1 (Z=0.766) M$_{\oplus}$,  when the heavies were represented by SiO$_2$ and H$_2$O , respectively.  These values are very similar to those obtained for Uranus. 
For Neptune, \citep{Movshovitz22} found that Neptune's interior consists of a relatively large amount of elements heavier than water. \cite{Neuenschwander2024} presented empirical models for Neptune and predicted the $J_6$ and $J_8$ values. It was also shown that that a precise measurement of the MoI of Neptune, with relative uncertainties of approximately 0.1\% (1\% for Uranus), could help to constrain the planetary rotation rate and depth of the winds. 
Like Uranus, Neptune could also have significant amounts of methane \citep{2024arXiv240312512M}.   
\newpage
\section{Summary \& Outlook}
Studying the interiors of outer planets  in the Solar  System, Jupiter, Saturn, Uranus, and Neptune, is vital for advancing our understanding of planetary formation and evolution as well as for the characterization of planets around other stars. 
In addition, these giant planets serve as natural laboratories for examining physical and chemical processes under extreme conditions that are not reproducible on Earth. 
Overall, our current knowledge of the outer planets can be summarized as follows:  
\begin{itemize}%
\item The internal structures and compositions of the outer planets  in the solar system must be inferred from numerical models that fit the available measurements. 
\item The planetary atmospheric composition provides an  important constraint for interior models but the link between the atmospheric composition and the bulk composition is non-trivial and requires further investigations. 
\item The composition of Jupiter and Saturn is dominated by hydrogen and helium but each planet has unique characteristics. 
\item The internal structure of giant planets is complex; the interior is not differentiated and includes composition gradients and “fuzzy” cores. Jupiter is less enriched with heavy elements compared to Saturn and has a more extended fuzzy core. In Saturn the process of helium rain is important, and above a heavy-element core a helium ocean may exist. 
\item The internal structures of Uranus and Neptune are not well understood. Both planets are expected to be composed of rocks and ices and have H-He atmospheres of the order of 10\% of their total masses.  The rock-to-water ratio in Uranus and Neptune is very uncertain. It is also unclear how the different materials are distributed within the interiors and whether distinct layers exist.
\item It remains to be determined how different the two planets from each other are, and whether Uranus and Neptune are indeed "icy". 
\item Advanced modeling, future observations from space and the ground, lab experiments, and links with exoplanetary science can improve our understanding of giant planets as a class of astronomical objects. 
\end{itemize}
\par

Despite great advancements in the last few decades, key open questions remain. However, the future of studying the interiors of outer planets looks bright  with advancements in both observational technologies and theoretical/numerical models.  Concurrently, refining our understanding of the equation of state for various elements and their interactions, while integrating diverse datasets (gravity fields, magnetic fields, atmospheric compositions, etc.), is crucial for refining our grasp of planetary interiors. Technological advancements in high-pressure physics experiments  and supercomputing will allow for an improved understanding of materials at extreme conditions.  In addition, future missions such as a Saturn atmospheric entry probe and orbiter+probe missions to Uranus and Neptune will provide key data on the atmospheric composition of the planets and their gravitational and magnetic fields, offering deeper insights into their bulk compositions and internal structures.
Finally, we are convinced that integrating information from different measurements and theoretical investigations can deepen our understanding of the fundamental processes governing the formation, evolution, interiors of the giant planets. 


\begin{ack}[Acknowledgments]
W We thank Guglielmo Mazzola and Luca Morf for support. We also acknowledge support from SNSF grant \texttt{\detokenize{200020_188460}} and the National Centre for Competence in Research ‘PlanetS’ supported by SNSF.
\end{ack}


\bibliographystyle{Harvard}
\begin{thebibliography*}{45}
\providecommand{\bibtype}[1]{}
\providecommand{\natexlab}[1]{#1}
{\catcode`\|=0\catcode`\#=12\catcode`\@=11\catcode`\\=12
|immediate|write|@auxout{\expandafter\ifx\csname
  natexlab\endcsname\relax\gdef\natexlab#1{#1}\fi}}
\renewcommand{\url}[1]{{\tt #1}}
\providecommand{\urlprefix}{URL }
\expandafter\ifx\csname urlstyle\endcsname\relax
  \providecommand{\doi}[1]{doi:\discretionary{}{}{}#1}\else
  \providecommand{\doi}{doi:\discretionary{}{}{}\begingroup
  \urlstyle{rm}\Url}\fi
\providecommand{\bibinfo}[2]{#2}
\providecommand{\eprint}[2][]{\url{#2}}

\bibtype{Article}%
\bibitem[{Bailey} and {Stevenson}(2021)]{Bailey2021}
\bibinfo{author}{{Bailey} E} and  \bibinfo{author}{{Stevenson} DJ}
  (\bibinfo{year}{2021}), \bibinfo{month}{Apr.}
\bibinfo{title}{{Thermodynamically Governed Interior Models of Uranus and
  Neptune}} \bibinfo{volume}{2} (\bibinfo{number}{2}), \bibinfo{eid}{64}.
  \bibinfo{doi}{\doi{10.3847/PSJ/abd1e0}}.
\eprint{2012.04166}.

\bibtype{Book}%
\bibitem[Beatty et al.(1999)]{beatty1999new}
\bibinfo{author}{Beatty J}, \bibinfo{author}{Petersen C} and
  \bibinfo{author}{Chaikin A} (\bibinfo{year}{1999}).
\bibinfo{title}{The New Solar System}, \bibinfo{publisher}{Cambridge University
  Press}.
\bibinfo{comment}{ISBN} \bibinfo{isbn}{9780521645874}.
\bibinfo{url}{\url{https://books.google.ch/books?id=iOezyHMVAMcC}}.

\bibtype{Article}%
\bibitem[{Debras} and {Chabrier}(2019)]{Debras2019}
\bibinfo{author}{{Debras} F} and  \bibinfo{author}{{Chabrier} G}
  (\bibinfo{year}{2019}), \bibinfo{month}{Feb.}
\bibinfo{title}{{New Models of Jupiter in the Context of Juno and Galileo}}.
\bibinfo{journal}{{\em \apj}} \bibinfo{volume}{872} (\bibinfo{number}{1}),
  \bibinfo{eid}{100}. \bibinfo{doi}{\doi{10.3847/1538-4357/aaff65}}.
\eprint{1901.05697}.

\bibtype{Article}%
\bibitem[{Fuller}(2014)]{Fuller2014}
\bibinfo{author}{{Fuller} J} (\bibinfo{year}{2014}), \bibinfo{month}{Nov.}
\bibinfo{title}{{Saturn ring seismology: Evidence for stable stratification in
  the deep interior of Saturn}}.
\bibinfo{journal}{{\em \icarus}} \bibinfo{volume}{242}:
  \bibinfo{pages}{283--296}. \bibinfo{doi}{\doi{10.1016/j.icarus.2014.08.006}}.
\eprint{1406.3343}.

\bibtype{Inproceedings}%
\bibitem[{Guillot} et al.(2023)]{guillot2023}
\bibinfo{author}{{Guillot} T}, \bibinfo{author}{{Fletcher} LN},
  \bibinfo{author}{{Helled} R}, \bibinfo{author}{{Ikoma} M},
  \bibinfo{author}{{Line} MR} and  \bibinfo{author}{{Paramentier} V}
  (\bibinfo{year}{2023}), \bibinfo{month}{Jul.}, \bibinfo{title}{{Giant Planets
  from the Inside-Out}}, \bibinfo{editor}{{Inutsuka} S},
  \bibinfo{editor}{{Aikawa} Y}, \bibinfo{editor}{{Muto} T},
  \bibinfo{editor}{{Tomida} K} and  \bibinfo{editor}{{Tamura} M}, (Eds.),
  \bibinfo{booktitle}{Protostars and Planets VII},
  \bibinfo{series}{Astronomical Society of the Pacific Conference Series},
  \bibinfo{volume}{534}, pp. \bibinfo{pages}{947}.

\bibtype{Article}%
\bibitem[{Helled} and {Fortney}(2020)]{HelledFortney2020}
\bibinfo{author}{{Helled} R} and  \bibinfo{author}{{Fortney} JJ}
  (\bibinfo{year}{2020}), \bibinfo{month}{Dec.}
\bibinfo{title}{{The interiors of Uranus and Neptune: current understanding and
  open questions}}.
\bibinfo{journal}{{\em Philosophical Transactions of the Royal Society of
  London Series A}} \bibinfo{volume}{378} (\bibinfo{number}{2187}),
  \bibinfo{eid}{20190474}. \bibinfo{doi}{\doi{10.1098/rsta.2019.0474}}.
\eprint{2007.10783}.

\bibtype{Article}%
\bibitem[{Helled} and {Stevenson}(2024)]{HelledAGU2024}
\bibinfo{author}{{Helled} R} and  \bibinfo{author}{{Stevenson} DJ}
  (\bibinfo{year}{2024}), \bibinfo{month}{Apr.}
\bibinfo{title}{{The Fuzzy Cores of Jupiter and Saturn}}.
\bibinfo{journal}{{\em AGU Advances}} \bibinfo{volume}{5}
  (\bibinfo{number}{2}), \bibinfo{eid}{e2024AV001171}.
  \bibinfo{doi}{\doi{10.1029/2024AV001171}}.
\eprint{2403.11657}.

\bibtype{Article}%
\bibitem[{Helled} et al.(2010)]{Helled2010shape}
\bibinfo{author}{{Helled} R}, \bibinfo{author}{{Anderson} JD} and
  \bibinfo{author}{{Schubert} G} (\bibinfo{year}{2010}), \bibinfo{month}{Nov.}
\bibinfo{title}{{Uranus and Neptune: Shape and rotation}}.
\bibinfo{journal}{{\em \icarus}} \bibinfo{volume}{210} (\bibinfo{number}{1}):
  \bibinfo{pages}{446--454}. \bibinfo{doi}{\doi{10.1016/j.icarus.2010.06.037}}.
\eprint{1006.3840}.

\bibtype{Article}%
\bibitem[{Helled} et al.(2011)]{helled11}
\bibinfo{author}{{Helled} R}, \bibinfo{author}{{Anderson} JD},
  \bibinfo{author}{{Podolak} M} and  \bibinfo{author}{{Schubert} G}
  (\bibinfo{year}{2011}), \bibinfo{month}{Jan.}
\bibinfo{title}{{Interior Models of Uranus and Neptune}}.
\bibinfo{journal}{{\em \apj}} \bibinfo{volume}{726} (\bibinfo{number}{1}),
  \bibinfo{eid}{15}. \bibinfo{doi}{\doi{10.1088/0004-637X/726/1/15}}.
\eprint{1010.5546}.

\bibtype{Article}%
\bibitem[{Helled} et al.(2020{\natexlab{a}})]{2020NatRP...2..562H}
\bibinfo{author}{{Helled} R}, \bibinfo{author}{{Mazzola} G} and
  \bibinfo{author}{{Redmer} R} (\bibinfo{year}{2020}{\natexlab{a}}),
  \bibinfo{month}{Sep.}
\bibinfo{title}{{Understanding dense hydrogen at planetary conditions}}.
\bibinfo{journal}{{\em Nature Reviews Physics}} \bibinfo{volume}{2}
  (\bibinfo{number}{10}): \bibinfo{pages}{562--574}.
  \bibinfo{doi}{\doi{10.1038/s42254-020-0223-3}}.
\eprint{2006.12219}.

\bibtype{Article}%
\bibitem[{Helled} et al.(2020{\natexlab{b}})]{Helled2020}
\bibinfo{author}{{Helled} R}, \bibinfo{author}{{Nettelmann} N} and
  \bibinfo{author}{{Guillot} T} (\bibinfo{year}{2020}{\natexlab{b}}),
  \bibinfo{month}{Mar.}
\bibinfo{title}{{Uranus and Neptune: Origin, Evolution and Internal
  Structure}}.
\bibinfo{journal}{{\em \ssr}} \bibinfo{volume}{216} (\bibinfo{number}{3}),
  \bibinfo{eid}{38}. \bibinfo{doi}{\doi{10.1007/s11214-020-00660-3}}.
\eprint{1909.04891}.

\bibtype{Article}%
\bibitem[{Helled} et al.(2022{\natexlab{a}})]{Helled2022b}
\bibinfo{author}{{Helled} R}, \bibinfo{author}{{Movshovitz} N} and
  \bibinfo{author}{{Nettelmann} N} (\bibinfo{year}{2022}{\natexlab{a}}),
  \bibinfo{month}{Feb.}
\bibinfo{title}{{The nature of gas giant planets}}.
\bibinfo{journal}{{\em arXiv e-prints}} ,
  \bibinfo{eid}{arXiv:2202.10046}\bibinfo{doi}{\doi{10.48550/arXiv.2202.10046}}.
\eprint{2202.10046}.

\bibtype{Article}%
\bibitem[{Helled} et al.(2022{\natexlab{b}})]{HelledStevenson2022}
\bibinfo{author}{{Helled} R}, \bibinfo{author}{{Stevenson} DJ},
  \bibinfo{author}{{Lunine} JI}, \bibinfo{author}{{Bolton} SJ},
  \bibinfo{author}{{Nettelmann} N}, \bibinfo{author}{{Atreya} S},
  \bibinfo{author}{{Guillot} T}, \bibinfo{author}{{Militzer} B},
  \bibinfo{author}{{Miguel} Y} and  \bibinfo{author}{{Hubbard} WB}
  (\bibinfo{year}{2022}{\natexlab{b}}), \bibinfo{month}{May}.
\bibinfo{title}{{Revelations on Jupiter's formation, evolution and interior:
  Challenges from Juno results}}.
\bibinfo{journal}{{\em \icarus}} \bibinfo{volume}{378}, \bibinfo{eid}{114937}.
  \bibinfo{doi}{\doi{10.1016/j.icarus.2022.114937}}.
\eprint{2202.10041}.

\bibtype{Article}%
\bibitem[{Howard} and {Guillot}(2023)]{hg23}
\bibinfo{author}{{Howard} S} and  \bibinfo{author}{{Guillot} T}
  (\bibinfo{year}{2023}), \bibinfo{month}{Apr.}
\bibinfo{title}{{Accounting for non-ideal mixing effects in the hydrogen-helium
  equation of state}}.
\bibinfo{journal}{{\em \aap}} \bibinfo{volume}{672}, \bibinfo{eid}{L1}.
  \bibinfo{doi}{\doi{10.1051/0004-6361/202244851}}.
\eprint{2302.07902}.

\bibtype{Article}%
\bibitem[{Howard} et al.(2023{\natexlab{a}})]{howard2023_interior}
\bibinfo{author}{{Howard} S}, \bibinfo{author}{{Guillot} T},
  \bibinfo{author}{{Bazot} M}, \bibinfo{author}{{Miguel} Y},
  \bibinfo{author}{{Stevenson} DJ}, \bibinfo{author}{{Galanti} E},
  \bibinfo{author}{{Kaspi} Y}, \bibinfo{author}{{Hubbard} WB},
  \bibinfo{author}{{Militzer} B}, \bibinfo{author}{{Helled} R},
  \bibinfo{author}{{Nettelmann} N}, \bibinfo{author}{{Idini} B} and
  \bibinfo{author}{{Bolton} S} (\bibinfo{year}{2023}{\natexlab{a}}),
  \bibinfo{month}{Apr.}
\bibinfo{title}{{Jupiter's interior from Juno: Equation-of-state uncertainties
  and dilute core extent}}.
\bibinfo{journal}{{\em \aap}} \bibinfo{volume}{672}, \bibinfo{eid}{A33}.
  \bibinfo{doi}{\doi{10.1051/0004-6361/202245625}}.
\eprint{2302.09082}.

\bibtype{Article}%
\bibitem[{Howard} et al.(2023{\natexlab{b}})]{howard2023_invZ}
\bibinfo{author}{{Howard} S}, \bibinfo{author}{{Guillot} T},
  \bibinfo{author}{{Markham} S}, \bibinfo{author}{{Helled} R},
  \bibinfo{author}{{M{\"u}ller} S}, \bibinfo{author}{{Stevenson} DJ},
  \bibinfo{author}{{Lunine} JI}, \bibinfo{author}{{Miguel} Y} and
  \bibinfo{author}{{Nettelmann} N} (\bibinfo{year}{2023}{\natexlab{b}}),
  \bibinfo{month}{Dec.}
\bibinfo{title}{{Exploring the hypothesis of an inverted Z gradient inside
  Jupiter}}.
\bibinfo{journal}{{\em \aap}} \bibinfo{volume}{680}, \bibinfo{eid}{L2}.
  \bibinfo{doi}{\doi{10.1051/0004-6361/202348129}}.
\eprint{2311.07646}.

\bibtype{Article}%
\bibitem[{Hubbard}(1969)]{hubbard1969}
\bibinfo{author}{{Hubbard} WB} (\bibinfo{year}{1969}), \bibinfo{month}{Jan.}
\bibinfo{title}{{Thermal Models of Jupiter and Saturn}}.
\bibinfo{journal}{{\em \apj}} \bibinfo{volume}{155}: \bibinfo{pages}{333}.
  \bibinfo{doi}{\doi{10.1086/149868}}.

\bibtype{Article}%
\bibitem[{Iess} et al.(2019)]{Iess2019}
\bibinfo{author}{{Iess} L}, \bibinfo{author}{{Militzer} B},
  \bibinfo{author}{{Kaspi} Y}, \bibinfo{author}{{Nicholson} P},
  \bibinfo{author}{{Durante} D}, \bibinfo{author}{{Racioppa} P},
  \bibinfo{author}{{Anabtawi} A}, \bibinfo{author}{{Galanti} E},
  \bibinfo{author}{{Hubbard} W}, \bibinfo{author}{{Mariani} MJ},
  \bibinfo{author}{{Tortora} P}, \bibinfo{author}{{Wahl} S} and
  \bibinfo{author}{{Zannoni} M} (\bibinfo{year}{2019}), \bibinfo{month}{Jun.}
\bibinfo{title}{{Measurement and implications of Saturn's gravity field and
  ring mass}}.
\bibinfo{journal}{{\em Science}} \bibinfo{volume}{364}
  (\bibinfo{number}{6445}), \bibinfo{eid}{aat2965}.
  \bibinfo{doi}{\doi{10.1126/science.aat2965}}.

\bibtype{Article}%
\bibitem[{Jeffreys}(1924)]{jeffreys1924}
\bibinfo{author}{{Jeffreys} H} (\bibinfo{year}{1924}), \bibinfo{month}{May}.
\bibinfo{title}{{On the internal constitution of Jupiter and Saturn}}.
\bibinfo{journal}{{\em \mnras}} \bibinfo{volume}{84}: \bibinfo{pages}{534}.
  \bibinfo{doi}{\doi{10.1093/mnras/84.7.534}}.

\bibtype{Article}%
\bibitem[{Leconte} and {Chabrier}(2012)]{Leconte2012}
\bibinfo{author}{{Leconte} J} and  \bibinfo{author}{{Chabrier} G}
  (\bibinfo{year}{2012}), \bibinfo{month}{Apr.}
\bibinfo{title}{{A new vision of giant planet interiors: Impact of double
  diffusive convection}}.
\bibinfo{journal}{{\em \aap}} \bibinfo{volume}{540}, \bibinfo{eid}{A20}.
  \bibinfo{doi}{\doi{10.1051/0004-6361/201117595}}.
\eprint{1201.4483}.

\bibtype{Article}%
\bibitem[{Malamud} et al.(2024)]{2024arXiv240312512M}
\bibinfo{author}{{Malamud} U}, \bibinfo{author}{{Podolak} M},
  \bibinfo{author}{{Podolak} J} and  \bibinfo{author}{{Bodenheimer} P}
  (\bibinfo{year}{2024}), \bibinfo{month}{Mar.}
\bibinfo{title}{{Uranus and Neptune as methane planets: producing icy giants
  from refractory planetesimals}}.
\bibinfo{journal}{{\em arXiv e-prints}} ,
  \bibinfo{eid}{arXiv:2403.12512}\bibinfo{doi}{\doi{10.48550/arXiv.2403.12512}}.
\eprint{2403.12512}.

\bibtype{Article}%
\bibitem[{Mankovich} and {Fortney}(2020)]{Mankovich2020}
\bibinfo{author}{{Mankovich} CR} and  \bibinfo{author}{{Fortney} JJ}
  (\bibinfo{year}{2020}), \bibinfo{month}{Jan.}
\bibinfo{title}{{Evidence for a Dichotomy in the Interior Structures of Jupiter
  and Saturn from Helium Phase Separation}}.
\bibinfo{journal}{{\em \apj}} \bibinfo{volume}{889} (\bibinfo{number}{1}),
  \bibinfo{eid}{51}. \bibinfo{doi}{\doi{10.3847/1538-4357/ab6210}}.
\eprint{1912.01009}.

\bibtype{Article}%
\bibitem[{Mankovich} and {Fuller}(2021)]{mankovich2021}
\bibinfo{author}{{Mankovich} CR} and  \bibinfo{author}{{Fuller} J}
  (\bibinfo{year}{2021}), \bibinfo{month}{Nov.}
\bibinfo{title}{{A diffuse core in Saturn revealed by ring seismology}}.
\bibinfo{journal}{{\em Nature Astronomy}} \bibinfo{volume}{5}:
  \bibinfo{pages}{1103--1109}. \bibinfo{doi}{\doi{10.1038/s41550-021-01448-3}}.
\eprint{2104.13385}.

\bibtype{Article}%
\bibitem[{Mazzola} et al.(2018)]{Mazzola2018}
\bibinfo{author}{{Mazzola} G}, \bibinfo{author}{{Helled} R} and
  \bibinfo{author}{{Sorella} S} (\bibinfo{year}{2018}), \bibinfo{month}{Jan.}
\bibinfo{title}{{Phase Diagram of Hydrogen and a Hydrogen-Helium Mixture at
  Planetary Conditions by Quantum Monte Carlo Simulations}}.
\bibinfo{journal}{{\em \prl}} \bibinfo{volume}{120} (\bibinfo{number}{2}),
  \bibinfo{eid}{025701}. \bibinfo{doi}{\doi{10.1103/PhysRevLett.120.025701}}.
\eprint{1709.08648}.

\bibtype{Article}%
\bibitem[{Miguel} et al.(2022)]{Miguel2022}
\bibinfo{author}{{Miguel} Y}, \bibinfo{author}{{Bazot} M},
  \bibinfo{author}{{Guillot} T}, \bibinfo{author}{{Howard} S},
  \bibinfo{author}{{Galanti} E}, \bibinfo{author}{{Kaspi} Y},
  \bibinfo{author}{{Hubbard} WB}, \bibinfo{author}{{Militzer} B},
  \bibinfo{author}{{Helled} R}, \bibinfo{author}{{Atreya} SK},
  \bibinfo{author}{{Connerney} JEP}, \bibinfo{author}{{Durante} D},
  \bibinfo{author}{{Kulowski} L}, \bibinfo{author}{{Lunine} JI},
  \bibinfo{author}{{Stevenson} D} and  \bibinfo{author}{{Bolton} S}
  (\bibinfo{year}{2022}), \bibinfo{month}{Jun.}
\bibinfo{title}{{Jupiter's inhomogeneous envelope}}.
\bibinfo{journal}{{\em \aap}} \bibinfo{volume}{662}, \bibinfo{eid}{A18}.
  \bibinfo{doi}{\doi{10.1051/0004-6361/202243207}}.
\eprint{2203.01866}.

\bibtype{Article}%
\bibitem[{Militzer} et al.(2019)]{militzer2019}
\bibinfo{author}{{Militzer} B}, \bibinfo{author}{{Wahl} S} and
  \bibinfo{author}{{Hubbard} WB} (\bibinfo{year}{2019}), \bibinfo{month}{Jul.}
\bibinfo{title}{{Models of Saturn's Interior Constructed with an Accelerated
  Concentric Maclaurin Spheroid Method}}.
\bibinfo{journal}{{\em \apj}} \bibinfo{volume}{879} (\bibinfo{number}{2}),
  \bibinfo{eid}{78}. \bibinfo{doi}{\doi{10.3847/1538-4357/ab23f0}}.
\eprint{1905.08907}.

\bibtype{Article}%
\bibitem[{Militzer} et al.(2022)]{Militzer2022}
\bibinfo{author}{{Militzer} B}, \bibinfo{author}{{Hubbard} WB},
  \bibinfo{author}{{Wahl} S}, \bibinfo{author}{{Lunine} JI},
  \bibinfo{author}{{Galanti} E}, \bibinfo{author}{{Kaspi} Y},
  \bibinfo{author}{{Miguel} Y}, \bibinfo{author}{{Guillot} T},
  \bibinfo{author}{{Moore} KM}, \bibinfo{author}{{Parisi} M},
  \bibinfo{author}{{Connerney} JEP}, \bibinfo{author}{{Helled} R},
  \bibinfo{author}{{Cao} H}, \bibinfo{author}{{Mankovich} C},
  \bibinfo{author}{{Stevenson} DJ}, \bibinfo{author}{{Park} RS},
  \bibinfo{author}{{Wong} M}, \bibinfo{author}{{Atreya} SK},
  \bibinfo{author}{{Anderson} J} and  \bibinfo{author}{{Bolton} SJ}
  (\bibinfo{year}{2022}), \bibinfo{month}{Aug.}
\bibinfo{title}{{Juno Spacecraft Measurements of Jupiter's Gravity Imply a
  Dilute Core}}.
\bibinfo{journal}{{\em \psj}} \bibinfo{volume}{3} (\bibinfo{number}{8}),
  \bibinfo{eid}{185}. \bibinfo{doi}{\doi{10.3847/PSJ/ac7ec8}}.

\bibtype{Article}%
\bibitem[{Morales} et al.(2013)]{Morales2013}
\bibinfo{author}{{Morales} MA}, \bibinfo{author}{{Hamel} S},
  \bibinfo{author}{{Caspersen} K} and  \bibinfo{author}{{Schwegler} E}
  (\bibinfo{year}{2013}), \bibinfo{month}{May}.
\bibinfo{title}{{Hydrogen-helium demixing from first principles: From diamond
  anvil cells to planetary interiors}}.
\bibinfo{journal}{{\em \prb}} \bibinfo{volume}{87} (\bibinfo{number}{17}),
  \bibinfo{eid}{174105}. \bibinfo{doi}{\doi{10.1103/PhysRevB.87.174105}}.

\bibtype{Article}%
\bibitem[Movshovitz and Fortney(2022)]{Movshovitz22}
\bibinfo{author}{Movshovitz N} and  \bibinfo{author}{Fortney J}
  (\bibinfo{year}{2022}).
\bibinfo{title}{{The Promise and Limitations of Precision Gravity: Application
  to the Interior Structure of Uranus and Neptune}}.
\bibinfo{journal}{{\em PSJ}} \bibinfo{volume}{3}: \bibinfo{pages}{88}.

\bibtype{Article}%
\bibitem[{Movshovitz} et al.(2020)]{2020ApJ...891..109M}
\bibinfo{author}{{Movshovitz} N}, \bibinfo{author}{{Fortney} JJ},
  \bibinfo{author}{{Mankovich} C}, \bibinfo{author}{{Thorngren} D} and
  \bibinfo{author}{{Helled} R} (\bibinfo{year}{2020}), \bibinfo{month}{Mar.}
\bibinfo{title}{{Saturn's Probable Interior: An Exploration of Saturn's
  Potential Interior Density Structures}}.
\bibinfo{journal}{{\em \apj}} \bibinfo{volume}{891} (\bibinfo{number}{2}),
  \bibinfo{eid}{109}. \bibinfo{doi}{\doi{10.3847/1538-4357/ab71ff}}.
\eprint{1912.02137}.

\bibtype{Article}%
\bibitem[{M{\"u}ller} and {Helled}(2024)]{2024ApJ...967....7M}
\bibinfo{author}{{M{\"u}ller} S} and  \bibinfo{author}{{Helled} R}
  (\bibinfo{year}{2024}), \bibinfo{month}{May}.
\bibinfo{title}{{Can Jupiter's Atmospheric Metallicity Be Different from the
  Deep Interior?}}
\bibinfo{journal}{{\em \apj}} \bibinfo{volume}{967} (\bibinfo{number}{1}),
  \bibinfo{eid}{7}. \bibinfo{doi}{\doi{10.3847/1538-4357/ad3738}}.
\eprint{2403.16273}.

\bibtype{Article}%
\bibitem[{Nettelmann} et al.(2013)]{nettel13}
\bibinfo{author}{{Nettelmann} N}, \bibinfo{author}{{Helled} R},
  \bibinfo{author}{{Fortney} JJ} and  \bibinfo{author}{{Redmer} R}
  (\bibinfo{year}{2013}), \bibinfo{month}{Mar.}
\bibinfo{title}{{New indication for a dichotomy in the interior structure of
  Uranus and Neptune from the application of modified shape and rotation
  data}}.
\bibinfo{journal}{{\em \planss}} \bibinfo{volume}{77}:
  \bibinfo{pages}{143--151}. \bibinfo{doi}{\doi{10.1016/j.pss.2012.06.019}}.
\eprint{1207.2309}.

\bibtype{Article}%
\bibitem[{Nettelmann} et al.(2021)]{nettelmann2021}
\bibinfo{author}{{Nettelmann} N}, \bibinfo{author}{{Movshovitz} N},
  \bibinfo{author}{{Ni} D}, \bibinfo{author}{{Fortney} JJ},
  \bibinfo{author}{{Galanti} E}, \bibinfo{author}{{Kaspi} Y},
  \bibinfo{author}{{Helled} R}, \bibinfo{author}{{Mankovich} CR} and
  \bibinfo{author}{{Bolton} S} (\bibinfo{year}{2021}), \bibinfo{month}{Dec.}
\bibinfo{title}{{Theory of Figures to the Seventh Order and the Interiors of
  Jupiter and Saturn}}.
\bibinfo{journal}{{\em \psj}} \bibinfo{volume}{2} (\bibinfo{number}{6}),
  \bibinfo{eid}{241}. \bibinfo{doi}{\doi{10.3847/PSJ/ac390a}}.
\eprint{2110.15452}.

\bibtype{Article}%
\bibitem[{Neuenschwander} and {Helled}(2022)]{Neuenschwander2022}
\bibinfo{author}{{Neuenschwander} BA} and  \bibinfo{author}{{Helled} R}
  (\bibinfo{year}{2022}), \bibinfo{month}{May}.
\bibinfo{title}{{Empirical structure models of Uranus and Neptune}}.
\bibinfo{journal}{{\em \mnras}} \bibinfo{volume}{512} (\bibinfo{number}{3}):
  \bibinfo{pages}{3124--3136}. \bibinfo{doi}{\doi{10.1093/mnras/stac628}}.
\eprint{2203.02233}.

\bibtype{Article}%
\bibitem[{Neuenschwander} et al.(2021)]{Neuenschwander2021}
\bibinfo{author}{{Neuenschwander} BA}, \bibinfo{author}{{Helled} R},
  \bibinfo{author}{{Movshovitz} N} and  \bibinfo{author}{{Fortney} JJ}
  (\bibinfo{year}{2021}), \bibinfo{month}{Mar.}
\bibinfo{title}{{Connecting the Gravity Field, Moment of Inertia, and Core
  Properties in Jupiter through Empirical Structural Models}}.
\bibinfo{journal}{{\em \apj}} \bibinfo{volume}{910} (\bibinfo{number}{1}),
  \bibinfo{eid}{38}. \bibinfo{doi}{\doi{10.3847/1538-4357/abdfd4}}.
\eprint{2101.12508}.

\bibtype{Article}%
\bibitem[{Neuenschwander} et al.(2024)]{Neuenschwander2024}
\bibinfo{author}{{Neuenschwander} BA}, \bibinfo{author}{{M{\"u}ller} S} and
  \bibinfo{author}{{Helled} R} (\bibinfo{year}{2024}), \bibinfo{month}{Jan.}
\bibinfo{title}{{Uranus' Complex Internal Structure}}.
\bibinfo{journal}{{\em arXiv e-prints}} ,
  \bibinfo{eid}{arXiv:2401.11769}\bibinfo{doi}{\doi{10.48550/arXiv.2401.11769}}.
\eprint{2401.11769}.

\bibtype{Article}%
\bibitem[{Ni}(2019)]{ni2019}
\bibinfo{author}{{Ni} D} (\bibinfo{year}{2019}), \bibinfo{month}{Dec.}
\bibinfo{title}{{Understanding Jupiter's deep interior: the effect of a dilute
  core}}.
\bibinfo{journal}{{\em \aap}} \bibinfo{volume}{632}, \bibinfo{eid}{A76}.
  \bibinfo{doi}{\doi{10.1051/0004-6361/201935938}}.

\bibtype{Article}%
\bibitem[{Ni}(2020)]{ni2020}
\bibinfo{author}{{Ni} D} (\bibinfo{year}{2020}), \bibinfo{month}{Jul.}
\bibinfo{title}{{Understanding Saturn's interior from the Cassini Grand Finale
  gravity measurements}}.
\bibinfo{journal}{{\em \aap}} \bibinfo{volume}{639}, \bibinfo{eid}{A10}.
  \bibinfo{doi}{\doi{10.1051/0004-6361/202038267}}.

\bibtype{Article}%
\bibitem[{Podolak} et al.(2019)]{Podolak19}
\bibinfo{author}{{Podolak} M}, \bibinfo{author}{{Helled} R} and
  \bibinfo{author}{{Schubert} G} (\bibinfo{year}{2019}), \bibinfo{month}{Aug.}
\bibinfo{title}{{Effect of non-adiabatic thermal profiles on the inferred
  compositions of Uranus and Neptune}}.
\bibinfo{journal}{{\em \mnras}} \bibinfo{volume}{487} (\bibinfo{number}{2}):
  \bibinfo{pages}{2653--2664}. \bibinfo{doi}{\doi{10.1093/mnras/stz1467}}.
\eprint{1905.09099}.

\bibtype{Article}%
\bibitem[{Podolak} et al.(2022)]{Podolak2022}
\bibinfo{author}{{Podolak} JI}, \bibinfo{author}{{Malamud} U} and
  \bibinfo{author}{{Podolak} M} (\bibinfo{year}{2022}), \bibinfo{month}{Aug.}
\bibinfo{title}{{Random models for exploring planet compositions I: Uranus as
  an example}} \bibinfo{volume}{382}, \bibinfo{eid}{115017}.
  \bibinfo{doi}{\doi{10.1016/j.icarus.2022.115017}}.
\eprint{2203.01139}.

\bibtype{Article}%
\bibitem[{Scheibe} et al.(2021)]{Scheibe2021}
\bibinfo{author}{{Scheibe} L}, \bibinfo{author}{{Nettelmann} N} and
  \bibinfo{author}{{Redmer} R} (\bibinfo{year}{2021}), \bibinfo{month}{Jun.}
\bibinfo{title}{{Thermal evolution of Uranus and Neptune. II. Deep thermal
  boundary layer}}.
\bibinfo{journal}{{\em \aap}} \bibinfo{volume}{650}, \bibinfo{eid}{A200}.
  \bibinfo{doi}{\doi{10.1051/0004-6361/202140663}}.
\eprint{2105.01359}.

\bibtype{Article}%
\bibitem[{Sch{\"o}ttler} and {Redmer}(2018)]{Schoettler2018}
\bibinfo{author}{{Sch{\"o}ttler} M} and  \bibinfo{author}{{Redmer} R}
  (\bibinfo{year}{2018}), \bibinfo{month}{Mar.}
\bibinfo{title}{{Ab Initio Calculation of the Miscibility Diagram for
  Hydrogen-Helium Mixtures}}.
\bibinfo{journal}{{\em \prl}} \bibinfo{volume}{120} (\bibinfo{number}{11}),
  \bibinfo{eid}{115703}. \bibinfo{doi}{\doi{10.1103/PhysRevLett.120.115703}}.

\bibtype{Article}%
\bibitem[{Stevenson} and {Salpeter}(1977)]{stevenson1977a}
\bibinfo{author}{{Stevenson} DJ} and  \bibinfo{author}{{Salpeter} EE}
  (\bibinfo{year}{1977}), \bibinfo{month}{Oct.}
\bibinfo{title}{{The phase diagram and transport properties for hydrogen-helium
  fluid planets.}}
\bibinfo{journal}{{\em \apjs}} \bibinfo{volume}{35}: \bibinfo{pages}{221--237}.
  \bibinfo{doi}{\doi{10.1086/190478}}.

\bibtype{Article}%
\bibitem[{Vazan} and {Helled}(2020)]{Vazan2020}
\bibinfo{author}{{Vazan} A} and  \bibinfo{author}{{Helled} R}
  (\bibinfo{year}{2020}), \bibinfo{month}{Jan.}
\bibinfo{title}{{Explaining the low luminosity of Uranus: a self-consistent
  thermal and structural evolution}}.
\bibinfo{journal}{{\em \aap}} \bibinfo{volume}{633}, \bibinfo{eid}{A50}.
  \bibinfo{doi}{\doi{10.1051/0004-6361/201936588}}.
\eprint{1908.10682}.

\bibtype{Article}%
\bibitem[{Wahl} et al.(2017)]{Wahl2017}
\bibinfo{author}{{Wahl} SM}, \bibinfo{author}{{Hubbard} WB},
  \bibinfo{author}{{Militzer} B}, \bibinfo{author}{{Guillot} T},
  \bibinfo{author}{{Miguel} Y}, \bibinfo{author}{{Movshovitz} N},
  \bibinfo{author}{{Kaspi} Y}, \bibinfo{author}{{Helled} R},
  \bibinfo{author}{{Reese} D}, \bibinfo{author}{{Galanti} E},
  \bibinfo{author}{{Levin} S}, \bibinfo{author}{{Connerney} JE} and
  \bibinfo{author}{{Bolton} SJ} (\bibinfo{year}{2017}), \bibinfo{month}{May}.
\bibinfo{title}{{Comparing Jupiter interior structure models to Juno gravity
  measurements and the role of a dilute core}}.
\bibinfo{journal}{{\em \grl}} \bibinfo{volume}{44} (\bibinfo{number}{10}):
  \bibinfo{pages}{4649--4659}. \bibinfo{doi}{\doi{10.1002/2017GL073160}}.
\eprint{1707.01997}.

\end{thebibliography*}

\end{document}